\documentclass[english,aip,reprint,manuscript=article,etalmode=truncate,maxauthors=0]{revtex4-1}

\usepackage{graphicx}
\usepackage{bm}
\usepackage{amsmath}
\usepackage[utf8]{inputenc}
\usepackage[T1]{fontenc}
\usepackage{mathptmx}
\usepackage{physics}
\usepackage{hyperref}
\usepackage{xcolor}
\usepackage{soul}

\begin{document}


\title{Total Angular Momentum Conservation in Ehrenfest Dynamics with a  Truncated Basis of  Adiabatic States} 



\author{Zhen Tao}
\email{taozhen@sas.upenn.edu}
\affiliation{Department of Chemistry, University of Pennsylvania, Philadelphia, Pennsylvania 19104, USA}
\author{Xuezhi Bian}
\affiliation{Department of Chemistry, University of Pennsylvania, Philadelphia, Pennsylvania 19104, USA}
\author{Yanze Wu}
\affiliation{Department of Chemistry, University of Pennsylvania, Philadelphia, Pennsylvania 19104, USA}
\author{Jonathan Rawlinson}
\affiliation{Department of Mathematics, University of Manchester, Manchester M13 9PL, UK}
\author{Robert G. Littlejohn}
\affiliation{Department of Physics, University of California, Berkeley, California 94720, USA}
\author{Joseph E. Subotnik}
\email{subotnik@sas.upenn.edu}
\affiliation{Department of Chemistry, University of Pennsylvania, Philadelphia, Pennsylvania 19104, USA}


\date{\today}

\begin{abstract}
    We show that standard Ehrenfest dynamics does not conserve linear and angular momentum when using a basis of truncated adiabatic states. However, we also show that previously proposed effective Ehrenfest equations of motion\cite{Amano2005,Krishna2007} involving the non-Abelian Berry force do maintain momentum conservation. As a numerical example, we investigate  the Kramers' doublet of the methoxy radical using generalized Hartree-Fock with spin-orbit coupling and confirm angular momentum is conserved with the proper equations of motion. Our work makes clear some of the limitations of the Born-Oppenheimer approximation when using {\em ab initio} electronic structure theory to treat systems with unpaired electronic spin degrees of freedom and we demonstrate that Ehrenfest dynamics can offer  much improved, qualitatively correct results.  
\end{abstract}

\maketitle 


\section{Introduction}

Given the speed and power of modern computational supercomputers, nonadiabatic dynamics is widely used today to study ultrafast photochemical and charge-transfer dynamics (as probed by state-of-the-art experiments).\cite{tretiak:2014:acr} When connecting with experiments, however, in practice  nonadiabatic simulations must almost always make a quantum-classical approximation\cite{Tully1998,Egorov1999,Kapral1999,Curchod2013,Romer2013,Crespo-Otero2018} in order to be computationaly feasible. The fundamental assumption of propagating nuclei classically and the electrons quantum mechanically inevitably raises the issue of how to correctly incorporate the feedback between the quantum and the classical subsystems. One of the most popular choices at present is the surface hopping approach, \cite{Tully1990,Barbatti2011,Wang2016,Subotnik2016} whereby a swarm of trajectories  move along a single adiabatic surface, and stochastically hop between surfaces to account for nonadiabatic effects.  As we have documented recently,\cite{Wu2022} the surface-hopping algorithm faces difficulties in the presence of spin-orbit coupling (SOC), which we will address in a separate paper.  A second approach, orthogonal to the surface-hopping ansatz, is to include the interaction between the quantum and classical subsystems in a mean-field way, which gives rise to the standard Ehrenfest approach\cite{Micha1983,Billing1983,Sawada1985,Li2005}--an approach which is always well-defined (with or without SOC).

The pros and cons of Ehrenfest dynamics are well-known within the community.\cite{Meyer1979,Horsfield2004,Parandekar2005,Parandekar2006,Curchod2018,Esch2020,Esch2021,Shu2023} Ehrenfest is most advantageous if there are frequent transitions between nearly parallel states, and one can work with either a handful or a dense manifold\cite{Fedorov2019} of such states. However, the mean-field approximation between the quantum and classical subsystems can break down when there is a strong coupling between the two subsystems (e.g. when two potential energy surfaces are very displaced from each other); in such a case, methods such as multi-configurational Ehrenfest have been developed.\cite{Shalashilin2009,Römer2013} Standard Ehrenfest also cannot account for decoherence nor achieve detailed balance,\cite{Parandekar2005,Parandekar2006} which are important when studying systems in the condensed phase; various correction schemes have been proposed to address these deficiencies as well\cite{Zhu2004,ZhuJ2004,Akimov2014,Esch2020,Esch2021,Shu2023}. 
As a sidenote, we mention that, in this paper, we will concern ourselves 
strictly with what is known as ``linear-response'' Ehrenfest, where the electronic wavefunction is expanded in a basis of adiabatic/diabatic states;
we will not concern ourselves with real-time Ehrenfest dynamics,\cite{Li2005,Isborn2007} whereby the electronic wavefunction is propagated directly in the original atomic orbital basis.

Now, as a practical matter, there are several means by which one can judge the value of any dynamical approach. Almost always, the very first constraint on any good dynamics algorithm is energy conservation; in fact, checking for energy conservation is usually the very first means of making sure one's code is free of bugs.\footnote{Energy conservation and frustrated hops are of course the reasons that Tully's algorithm approximately obeys detailed balance.} Beyond energy conservation, 
in the absence of an external torque on the system, a second constraint is that the total angular momentum is also conserved. Interestingly, as opposed to energy conservation, angular momentum conservation is rarely emphasized in the context of nonadiabatic dynamics (though see below). Consider the most standard class of molecular dynamics, namely single-state dynamics within the Born-Oppenheimer (BO) approximation. For such dynamics, the electronic linear and angular momentum are usually neglected (or more formally folded into the nuclear degrees of freedom),\cite{LittleJohn2023}  and the {\em nuclear} linear and angular momentum are conserved due to translational symmetry and isotropy of space. However, as we recently showed,\cite{Bian2023} if one runs BO dynamics (without Berry force) along one of the degenerate doublet surfaces (in the presence of spin-orbit coupling) and keeps track of the  fluctuating electronic angular momentum, one will inevitably compute a total angular momentum that fluctuates (i.e. one will predict a violation of angular momentum conservation). 

Now, one means to resolve this paradox is not to use Born-Oppenheimer at all, but rather exact factorization,\cite{Abedi2010,Abedi2012,Vindel-Zandbergen2022} where angular momentum exchange was recently explored.\cite{gross:2022:prl:angmom} As pointed out in Ref. \citenum{Bian2023}, however, an even simpler resolution to this paradox is that, within BO theory, one must include a Berry force (see below) acting on the nuclear motion. More precisely, 
for nonadiabatic systems with odd numbers of electrons plus SOC, the on-diagonal derivative coupling $\bm d$ is not zero, and so the
nuclear kinetic momentum $\bm \pi_{\rm n}$ is not equivalent to the canonical momentum $\bm P_{\rm n}$ (recall that $\bm \pi_{\rm n} = \bm P_{\rm n} - i\hbar\bm d$).  In order to minimize gauge problems, the usual approach is then to pick one adiabat from the Kramers' pair to run along (say, $j$, which is computed from some approximate electronic structure technique) and then to propagate the nuclear kinetic momentum $\bm \pi_{\rm n}$; the latter step inevitably introduces the Abelian Berry force in the equation of motion for $\bm \pi_{\rm n}$,  $\bm F^{{\rm Berry},I}_{j} = i\hbar (\nabla_{\rm n}\cross\bm{d}_{jj})\cdot\frac{\bm \pi_{\rm n}}{M_{I}}$.
As shown  in Ref.\citenum{Bian2023}, including the pseudo-magnetic Berry force allows for a full exchange of angular momentum between electronic, nuclear, and spin degrees of freedom.
Of course, there is still no guarantee that the dynamics are correct (i.e. following one adiabat of a pair), but at least the resulting dynamics are guaranteed to conserve the total angular momentum.

The background above raises crucial questions for the field of nonadiabatic dynamics. If one needs to go beyond BO dynamics, one can ask: do nonadiabatic dynamics algorithms conserve the total (electronic plus nuclear) momentum? In a recent paper,  Shu et. al\cite{Shu2020} have recently shown that the nuclear angular momentum is not conserved within an {\em ab initio} Ehrenfest scheme propagated in an adiabatic basis. 
In Ref. \citenum{Shu2020}, the authors addressed this issue by projecting out the translational and rotational components of the derivative coupling that enters the force; see Eq. \ref{eqn:Pn-dot-adiab}. While this scheme offers a practical way to conserve nuclear angular momentum, we will show below that the problems arising in Ref.~\citenum{Shu2020} are at bottom created by using a truncated adiabatic basis, for which there is a rigorous (not {\em ad hoc}) solution.  Deriving and understanding such a  solution is the main focus of the present paper but in a nutshell, if BO dynamics require the Abelian Berry curvature in order to maintain momentum conservation, nonadiabatic Ehrenfest dynamics in a truncated basis require the non-Abelian Berry curvature \cite{mead:1992:rmp} in order to achieve the same feat.

Finally, before concluding this section, a few words are appropriate regarding spin. The most obvious cases where we expect angular momentum conservation to be interesting, are systems with a flow of angular momentum between different degrees of freedom (including  nuclear, electronic and spin degrees of freedom). For organic systems, the spin degree of freedom often operates on a much longer time scale than the electronic degree of freedom and sometimes even longer than the nuclear motion. In such cases, the validity of the BO approximation is dubious.
Indeed, in this paper we will show that a simple rotation of the methoxy radical breaks the BO approximation because, within the BO approximation, the total spin vector rotates with the molecule instantaneously (which is incorrect).  One would hope that the Ehrenfest equations of motion would perform far better, and indeed Ehrenfest does (correctly) slows down the spin change with the nuclear motion.

This paper is organized as follows. In Section \ref{sec:theory}, we begin by demonstrating  momentum conservation for Ehrenfest dynamics propagated over a complete electronic Hilbert space; this conclusion holds whether one performs the dynamics in a diabatic or adiabatic basis, and by comparing the calculations in the two different basis sets, one inevitably learns about the non-Abelian Berry curvature. In Sec. \ref{sec:adiab}, we then remove the assumption of a complete set of states and show that, according to standard Ehrenfest dynamics, neither linear and angular momentum are conserved in a truncated set of states. To restore momentum conservation in the presence of a truncated basis, we show that  the equations of motion must include the non-Abelian Berry curvature and we present the relevant Hamiltonian from which Hamilton's equations can be derived (where the final form agrees with the derivations in Refs. \citenum{Amano2005} and \citenum{Krishna2007}). In Sec. \ref{sec:results}, in order to demonstrate the importance of momentum conservation, we  perform two {\em ab initio} Ehrenfest calculations of the methoxy radical in the Kramers' pair basis. We study both excitation of a vibration and excitation of angular momentum. These two cases make  clear that including the non-Abelian Berry curvature can have a strong impact on the resulting spin dynamics and that, more generally, the BO approximation can badly break down in the presence of unpaired electrons.
In Sec.\ref{sec:conclusion}, we conclude and discuss future possible {\em ab initio} directions.

\section{Theory: Dynamics Within A Complete (Untruncated) Electronic Hilbert Space}\label{sec:theory}

We begin our analysis by assuming that we are working in a complete (untruncated) electronic vector space with zero curvature.  This scenario represents a very ideal condition because the electronic Hilbert space is immense -- requiring an enormous number (infinite) of one-particle electronic basis functions and then an even larger number (infinite) of many-body electronic wavefunctions (and just about any finite basis will exhibit a nonzero Berry curvature\cite{mead:1992:rmp}).  Nevertheless, the analysis below will still be useful insofar as teaching us how to understand how the Ehrenfest equations can take different form in different representations.

\subsection{A Strictly Diabatic Representation} \label{subsec:diab}
To begin our discussion, let us imagine that we are given an electronic Hamiltonian expressed in a strictly diabatic basis; in other words, the electronic basis does not depend at all on nuclear position.  The Hamiltonian ($\tilde{ \bm H}$) and the energy ($\tilde{E}$) are postulated to be of the form:
\begin{eqnarray}
    \label{eqn:H_diab}
      \tilde{\bm H} &=& \sum_{I}\frac{{ \bm P}_{\rm n}^{I^2}}{2M_{I}}+  \tilde{\bm V}\\
      \label{e_diab}
\tilde{E} &=&\sum_{I}\frac{{ \bm P}_{\rm n}^{I^2}}{2M_{I}}+\mbox{Tr}\left(\tilde{\bm \sigma}\tilde{\bm V}\right)
\end{eqnarray}
where we denote the classical nuclear position $\bm R_{\rm n}$ and nuclear momentum $\bm P_{\rm n}$. 
Here and below we use the indices $IJ$ for nuclei and $\alpha\beta\gamma$ for the Cartesian indices $xyz$.  The potential operator  $\tilde{\bm V}= \tilde{\bm T}_{\rm e} + \tilde{\bm V}_{\rm ee} + \tilde{\bm V}_{\rm en} + \tilde{\bm V}_{\rm nn}$ includes the electronic kinetic energy, electron-electron interaction, and the electron-nuclear Coulomb interaction, and nuclear-nuclear repulsion term, respectively.  We use the notation  "\char`\~" to indicate operators in a diabatic basis.  For the energy expression, the first term is the nuclear kinetic energy and the second term is the potential energy term that one computes by integrating over the electronic degree of freedom with the electronic density operator  $\tilde{\bm \sigma}$  in a diabatic basis. 

According to  Hamilton's equations, the equations of motions for nuclear position and momentum are:
\begin{eqnarray}
\label{eqn:R-dot-diab}
\dot{R}^{I\alpha}_{\rm n} &=& \frac{\partial \tilde{E}}{\partial P^{I\alpha}_{\rm n}} = \frac{P^{I\alpha}_{\rm n}}{M_{I}} \\
\label{eqn:P-dot-diab}
 \dot{P}^{I\alpha}_{\rm n} &=& -\frac{\partial \tilde{E}}{\partial R^{I\alpha}_{\rm n}} =   -\mbox{Tr}\left(\tilde{\bm \sigma} \frac{\partial \tilde{\bm V}}{\partial R^{I\alpha}_{\rm n}}\right) 
\end{eqnarray}
The associated density matrix operator evolves according to the Quantum Liouville equation,
\begin{equation}
\label{eqn: sigma-dot-diab}
\dot{\tilde {\bm \sigma}} =  - \frac{i}{\hbar} \left[ \tilde{\bm V}, \tilde{\bm \sigma}  \right] 
\end{equation}
Using Eqs. \ref{eqn:R-dot-diab} and \ref{eqn:P-dot-diab}, it is straightforward to show that the total energy in diabatic representation (Eq. \ref{e_diab}) is conserved $\frac{d\tilde{E}}{dt} = 0 $. 

At this point, it will be helpful to define nuclear angular momentum:
\begin{align}
\label{eqn:Jn1}
    {J}^{\alpha}_{\rm n} \equiv \sum_{I\beta\gamma} \epsilon_{\alpha\beta\gamma}  R^{I\beta}_{\rm n} M_I \dot{R}^{I\gamma}_{\rm n} 
\end{align}
\noindent Here $\epsilon_{\alpha\beta\gamma}$ is the Levi-Civita symbol. Using Eq. \ref{eqn:R-dot-diab} above, it follows that we can also write:
\begin{align}
    \label{eqn:Jn2}
    {J}^{\alpha}_{\rm n} = \sum_{I\beta\gamma} \epsilon_{\alpha\beta\gamma}  R^{I\beta}_{\rm n} P^{I\gamma}_{\rm n} \\ 
    \dot{J}^{\alpha}_{\rm n} = \sum_{I\beta\gamma} \epsilon_{\alpha\beta\gamma}  R^{I\beta}_{\rm n} \dot{P}^{I\gamma}_{\rm n} \label{eqn:Jn3}
\end{align}
\noindent With Eq.\ref{eqn:P-dot-diab} and Eq.\ref{eqn: sigma-dot-diab}, we can now evaluate the time-derivative of the total linear and angular momentum
\begin{eqnarray}
\label{eqn:tot_Pmom}
    \dot{ P}_{\rm tot}^{\alpha}  &=& \dot{ P}_{\rm n}^{\alpha} + \mbox{Tr}\left( \dot{\tilde{\bm \sigma}}  {{\tilde{\bm P}}}_{\rm e}^{\alpha} \right) \nonumber\\&=& -\sum_{I}\mbox{Tr}\left(\tilde{\bm \sigma} \frac{\partial \tilde{\bm V}}{\partial R^{I\alpha}_{\rm n}}\right) -\frac{i}{\hbar} \mbox{Tr}\left( \tilde{\bm \sigma}[{{\tilde{\bm P}}}_{\rm e}^{\alpha}, \tilde{\bm V}]\right)\\
   \dot{ J}_{\rm  tot}^{\alpha}  &=& \dot{ J}_{\rm n}^{\alpha} + \mbox{Tr}\left( \dot{\tilde{\bm \sigma}}  {{\tilde{\bm J}}}_{\rm e}^{\alpha} \right)\nonumber \\
   &=& \sum_{I\beta\gamma}\epsilon_{\alpha\beta\gamma}R^{I\beta}_{\rm n}\dot{P}^{I\gamma}_{\rm n} -\frac{i}{\hbar} \mbox{Tr}\left( \tilde{\bm \sigma}[{{\tilde{\bm J}}}_{\rm e}^{\alpha}, \tilde{\bm V}]\right)\label{eqn:tot_Jmom}
\end{eqnarray}

 Finally, {\em because we have assumed a complete electronic Hilbert space}, for a finite system in real space, translational symmetry and the isotropy of space imply the following identities:
\begin{eqnarray}
\label{eq:pcommutator}
    \Big[\tilde{\bm P}_{\rm n} + \tilde{\bm P}_{\rm e}, \tilde{\bm V}\Big] &=& 0 \\
    \Big[\tilde{\bm J}_{\rm n} + \tilde{\bm J}_{\rm e}, \tilde{\bm V}\Big] &=& 0 \label{eq:jcommutator}
\end{eqnarray}
These equations lead to the following further identities:
\begin{eqnarray}
    -\frac{i}{\hbar} [{{\tilde{\bm P}}}_{\rm e}^{\alpha}, \tilde{\bm V}] &=& \frac{i}{\hbar} [{{\tilde{\bm P}}}_{\rm n}^{\alpha}, \tilde{\bm V}] =\frac{\partial \tilde{\bm V}}{\partial R^{\alpha}_{\rm n} }\\
    -\frac{i}{\hbar} [{{\tilde{\bm J}}}_{\rm e}^{\alpha}, \tilde{\bm V}] &=&\frac{i}{\hbar} [{{\tilde{\bm J}}}_{\rm n}^{\alpha}, \tilde{\bm V}] =  \sum_{I\beta\gamma}\epsilon_{\alpha\beta\gamma}R^{I\beta}_{\rm n} \frac{\partial \tilde{\bm V}}{\partial R^{I\gamma}_{\rm n} }
\end{eqnarray}
If we plug the commutators above into Eq. \ref{eqn:tot_Pmom} and Eq.\ref{eqn:tot_Jmom}, we find momentum conservation $\dot{ P}_{\rm tot}^{\alpha}  = \dot{ J}_{\rm tot}^{\alpha} =0$.

\subsection{Adiabatic Representation} \label{subsec:adiab}
The above equations of motion for Ehrenfest dynamics can be transformed into an adiabatic basis as well with the same conclusions, though we will find that the existence of a complete electronic basis is expressed differently than what we found in Eqs. \ref{eq:pcommutator}-\ref{eq:jcommutator}.
To proceed,  let us define a unitary matrix that transforms the diabatic basis (with indices $abcd$) to adiabatic basis (with indices $ijkl$): $\ket{\psi_{k}}=\sum_{a} \ket{\phi_{a}} U_{ak}$. The density and potential operators in the adiabatic basis obtained after the diabatic-to-adiabatic transformation are,\begin{eqnarray}
    \hat{\bm \sigma} &= &\bm U^{\dagger}\tilde{\bm \sigma} \bm U \\
        \hat{\bm V} &=&\bm U^{\dagger} \tilde{\bm V} \bm U
\end{eqnarray}
To transform the equations of motions in the diabatic basis (Eq.\ref{eqn:R-dot-diab}-\ref{eqn: sigma-dot-diab}) to the adiabatic basis, let us write the equations of motion in terms of $\hat{\bm \sigma} $ and $\hat{\bm V} $. Specifically, 
\begin{eqnarray}
        \label{eqn:R-dot-adiab}
\dot{R}^{I\alpha}_{\rm n} &=& \frac{P^{I\alpha}_{\rm n}}{M_I} \\
        \dot{P}^{I\alpha}_{\rm n} &=&-\mbox{Tr}\left(\bm U^{\dagger} \tilde{\sigma} \bm U \bm U^{\dagger} \frac{\partial \tilde {\bm V} }{\partial R^{I\alpha}_{\rm n}}\bm U \right) \nonumber \\
        &=& -\mbox{Tr}\left(\hat{\sigma} \frac{\partial \hat{\bm V}}{\partial R^{I\alpha}_{\rm n}}\right) + \frac{i}{\hbar}
 \mbox{Tr}\left(\hat{\bm \sigma} \left[  {\bm A}^{I\alpha}, \hat{\bm V} \right] \right)\label{eqn:Pn-dot-adiab} \\
 \dot{\hat{\bm \sigma}} &=& \frac{d}{dt} \left(\bm U^{\dagger} \tilde{\bm \sigma} \bm U\right) = -\frac{i}{\hbar} \left[ \hat{\bm V} - \sum_{I}
\frac{{\bm P_{\rm n}^{I}} \cdot  {{\bm A}^{I}} }{M_{I}}, \hat{\bm \sigma} \right]  \label{eqn:d-dot-adiab}
\end{eqnarray}
Here we have defined 
\begin{equation}
    A^{I\alpha}_{jk} = i\hbar  \sum_{a} U^\dagger_{ja}  \frac{\partial U_{ak}} {\partial R^{I\alpha}_{\rm n}} = i\hbar\bra{\psi_{j}}\ket{\frac{\partial \psi_{k}}{\partial R^{I\alpha}_{\rm n}}},
\end{equation}
which is commonly known as the nonadiabatic coupling term or the Berry connection. It is also the negative of the nuclear momentum operator.
\begin{equation}
    \label{eqn:AandPn}
     A^{I\alpha}_{jk} = -\bra{\psi_j}\hat{ P}_{\rm n}^{I\alpha} \ket{\psi_k}
\end{equation}
Eqs.\ref{eqn:R-dot-adiab}-\ref{eqn:d-dot-adiab} are often considered the standard Ehrenfest equations of motion in an adiabatic basis.\cite{doltsinis:2002:review,Curchod2013}

To demonstrate momentum conservation within this adiabatic representation, we again evaluate the time-derivative of the total linear and angular momentum.
\begin{eqnarray}
\label{eqn:Ptot_dot}
\dot { P}_{\rm tot}^{\alpha} &=&  \sum_I \dot {P}_{\rm n}^{I\alpha}  +  \mbox{Tr}\left( \hat{\bm \sigma}  \dot {\bm P}_{\rm e}^{\alpha}  + \dot {\hat{\bm \sigma}}   {\bm P}_{\rm e}^{\alpha} \right) 
\\
\dot { J}_{\rm tot}^{\alpha} &=&   \sum_I \dot {J}_{\rm n}^{I\alpha}  +   \mbox{Tr}\left( \hat{\bm\sigma}  \dot {\bm J}_{\rm e}^{\alpha}  + \dot {\hat{\bm \sigma}}   {\bm J}_{\rm e}^{\alpha} \right)
\label{eqn:Jtot_dot}
\end{eqnarray}
Note that when propagating the equations of motion in the adiabatic basis, one can choose an arbitrary phase of the adiabatic state as long as it is smooth in the configuration space. For instance, let us assume that, in the vicinity of configuration $\bm R_0$, the electronic state is chosen as 
\begin{equation}
    \psi_{k}(\bm r;\bm R_{\rm n}) = \phi_k(\bm r-\bm R_{\rm n})e^{\frac{i}{\hbar} \zeta_{k}(\bm R_{\rm n}-\bm R_{0})}
\end{equation} 
In such a case, one finds the relations below:
\begin{eqnarray}
\label{eqn:sum_P}
 \Big(\sum_{I}\hat { P}^{I\alpha}_{\rm n} +  \hat { P}^{\alpha}_{\rm e}\Big)\ket{\psi_{k}} &=&  \sum_{I}\xi^{I\alpha}_{k}(\bm R_{\rm n})\ket{\psi_{k}}   \\
 \Big(\sum_{I}\hat { J}^{I\alpha}_{\rm n}+  \hat { J}^{\alpha}_{\rm e}\Big)\ket{\psi_{k}} &=&   \sum_{I\eta\gamma} \epsilon_{\alpha\eta\gamma} R^{I\eta}_{\rm n}\xi^{I\gamma}_{k}(\bm R_{\rm n})\ket{\psi_{k}} 
 \label{eqn:sum_J}
\end{eqnarray}
where $\xi^{I\alpha}_{k} (\bm R_{\rm n})=\nabla_{I\alpha}\zeta_{k} (\bm R_{\rm n})$.

Thereafter, one can arrive at the following identities for the matrix elements of the electronic momentum and angular momentum operators:
\begin{eqnarray}
    \label{eqn:AandP}
    \bra{\psi_j} \hat{ P}^{\alpha}_{\rm e} \ket{\psi_k}  &=& - \sum_{I}\bra{\psi_j} \hat{ P}^{I\alpha}_{\rm n} \ket{\psi_k} +\sum_{I}\xi^{I\alpha}_{k}\delta_{jk} \nonumber\\
    &=& \sum_{I}A^{I\alpha}_{jk} +\xi^{I\alpha}_{k}\delta_{jk} \\
    \label{eqn:AandJ}
    \bra{\psi_j} \hat{ J}^{\alpha}_{\rm e} \ket{\psi_k}  &=& - \sum_{I}\bra{\psi_j} \hat{ J}^{I\alpha}_{\rm n} \ket{\psi_k} +\sum_{I\eta\gamma} \epsilon_{\alpha\eta\gamma}R^{I\eta}_{\rm n}\xi^{I\gamma}_{k}\delta_{jk}\nonumber \\
    &=&  \sum_{I\eta\gamma} \epsilon_{\alpha\eta\gamma}R^{I\eta}_{\rm n}(A^{I\gamma}_{jk}+\xi^{I\gamma}_{k}\delta_{jk})
\end{eqnarray}
If we differentiate the matrix elements above with respect to time, we find:
\begin{eqnarray}
    \label{eqn:AandP_dot}
    \dot{ P}^{\alpha}_{{\rm e},jk}  &=&  \sum_{I}\dot{A}^{I\alpha}_{jk} +\dot{\xi}^{I\alpha}_{k}\delta_{jk}\nonumber \\
    &=&   \sum_{IJ\beta} \dot{R}^{J\beta}_{\rm n}\Bigg(\frac{\partial A^{I\alpha}_{jk}}{\partial R^{J\beta}_{\rm n}} +\xi^{J\beta,I\alpha}_{k}\delta_{jk}\Bigg)\\
    \label{eqn:AandJ_dot}
    \dot{ J}^{\alpha}_{{\rm e},jk}  &=&  \sum_{I\eta\gamma} \epsilon_{\alpha\eta\gamma}\dot{R}^{I\eta}_{\rm n}(A^{I\gamma}_{jk}+\xi^{I\gamma}_{k}\delta_{jk}) \nonumber
    \\ && +  \sum_{IJ\eta\gamma\eta} \epsilon_{\alpha\eta\gamma}R^{I\eta}_{\rm n} \dot{R}^{J\beta}_{\rm n}\Bigg(\frac{\partial A^{I\gamma}_{jk}}{\partial R^{J\beta}_{\rm n}} +\xi^{J\beta,I\gamma}_{k}\delta_{jk}\Bigg)
\end{eqnarray}
Here we introduce the notation $\xi^{J\beta,I\alpha}_{k}(\bm R_{\rm n}) = \nabla_{J\beta}\xi^{I\alpha}_{k}(\bm R_{\rm n})$.

\subsubsection{Linear Momentum Conservation}
To demonstrate conservation of linear momentum, let us now evaluate all terms in Eq. \ref{eqn:Ptot_dot}

\begin{itemize}
\item From Eq. \ref{eqn:Pn-dot-adiab}, the first term in Eq. \ref{eqn:Ptot_dot}  ($\sum_I \dot {P}_{\rm n}^{I\alpha}$) is 
\begin{equation}
\label{eqn:Pn-dot-adiab2}
\begin{split}
     \sum_I \dot {P}_{\rm n}^{I\alpha}
    =& \sum_I -\mbox{Tr}\left(\hat{\bm\sigma} \frac{\partial \hat{\bm V}}{\partial R^{I\alpha}_{\rm n}}\right) + \frac{i}{\hbar}
 \mbox{Tr}\left(\hat{\bm \sigma} \left[  {\bm A}^{I\alpha}, \hat{\bm V} \right] \right) \\
 =& \frac{i}{\hbar}\sum_I 
 \mbox{Tr}\left(\left[ \hat{\bm V}  ,\hat{\bm \sigma}   \right] {\bm A}^{I\alpha}\right) ,
\end{split}
\end{equation}
where the first term above vanishes since we assume a translationally invariant potential energy surface.
\item From Eq. \ref{eqn:AandP_dot}, the second term in Eq. \ref{eqn:Ptot_dot} [$\mbox{Tr}\left( \hat{\bm \sigma}  \dot {\bm P}_{\rm e}^{\alpha}\right)$] is
\begin{eqnarray}
   \label{eqn:sigma_Pe_dot}
\mbox{Tr}\left( \hat{\bm \sigma}  \dot {\bm P}_{\rm e}^{\alpha}\right) = \sum_{IJ\beta jk}\hat{\sigma}_{kj} \dot{R}^{J\beta}_{\rm n}\Bigg(\frac{\partial A^{I\alpha}_{jk}}{\partial R^{J\beta}_{\rm n}} +\xi^{J\beta,I\alpha}_{k}\delta_{jk}\Bigg)    \\ \nonumber
\end{eqnarray}

\item Using Eq.\ref{eqn:AandP} to express $\bm P^{\alpha}_{\rm e}$ in terms of $\bm A^{I\alpha}$, the last term  [$\mbox{Tr}\left(\dot {\hat{\bm \sigma}}   {\bm P}_{\rm e}^{\alpha}\right)$] in Eq.\ref{eqn:Ptot_dot} becomes
\begin{equation}
\label{eqn:d_dot_Pe_1}
\begin{split}
\mbox{Tr}\left(\dot {\hat{\bm \sigma}} {\bm P}_{\rm e}^{\alpha}\right) &= \sum_{Ijk}\dot{\hat{\bm \sigma}}_{kj}\left(A^{I\alpha}_{jk} +\xi^{I\alpha}_{k}\delta_{jk}\right) 
\end{split}
\end{equation}
If we plug Eq. \ref{eqn:d-dot-adiab} into Eq.\ref{eqn:d_dot_Pe_1}, the second term in Eq.\ref{eqn:d_dot_Pe_1} becomes,
\begin{eqnarray}
     \sum_{Ijk}\dot{\hat{\bm \sigma}}_{kj}\xi^{I\alpha}_{k}\delta_{jk}
    &=& -\frac{i}{\hbar}\sum_{Ijk}\xi^{I\alpha}_{k}\Big(\hat{V}_{kk}\hat{\sigma}_{kj}- \hat{\sigma}_{kj}\hat{V}_{jj}\Big)\delta_{jk}\nonumber\\&& +\frac{i}{\hbar}\sum_{IJ\beta jk} \dot{R}^{J\beta}_{\rm n} \hat{\sigma}_{kj }(\xi^{I\alpha}_{j}-\xi^{I\alpha}_{k})  A^{J\beta}_{jk} \\
    &=& - \frac{i}{\hbar}  \sum_{IJ\beta jk} \dot{R}^{J\beta}_{\rm n} \hat{\sigma}_{kj}(\xi^{I\alpha}_{k}-\xi^{I\alpha}_{j})  A^{J\beta}_{jk} \label{eqn:d_dot_Pe_2}
\end{eqnarray}
If we plug Eqs. \ref{eqn:d-dot-adiab} and \ref{eqn:d_dot_Pe_2} into  Eq.\ref{eqn:d_dot_Pe_1}, w can then simplify the total expression in Eq.\ref{eqn:d_dot_Pe_1}
\begin{equation}
\label{eqn:d_dot_Pe}
\begin{split}
     \mbox{Tr}\left(\dot {\hat{\bm \sigma}}   {\bm P}_{\rm e}^{\alpha}\right)
 = &-\frac{i}{\hbar} \sum_{I}
 \mbox{Tr}\left(\left[ \hat{\bm V}  ,\hat{\bm \sigma}   \right] {\bm A}^{I\alpha}\right)\\
 & +\frac i \hbar \mbox{Tr}\left( \hat{\bm \sigma}\sum_{IJ\beta} \dot R^{J\beta}_{\rm n} \left[ {\bm A}^{I\alpha},{\bm A}^{J\beta} \right]\right)  \\
 & -\frac{i}{\hbar} \sum_{IJ\beta jk} \dot{R}^{J\beta}_{\rm n} \hat{\sigma}_{kj }(\xi^{I\alpha}_{k}-\xi^{I\alpha}_{j})  A^{J\beta}_{jk} 
\end{split}
\end{equation}
\end{itemize}

Using Eqs. \ref{eqn:Pn-dot-adiab2},  \ref{eqn:sigma_Pe_dot}, and \ref{eqn:d_dot_Pe}, we can finally evaluate the time-dependence of the total linear momentum in Eq.\ref{eqn:Ptot_dot}. We notice that Eq. \ref{eqn:Pn-dot-adiab2} cancels with the first term in Eq. \ref{eqn:d_dot_Pe}, and we are left with
\begin{widetext}
\begin{eqnarray}
        \dot {P}_{\rm tot}^{\alpha} 
    &=& \sum_{IJ\beta} \mbox{Tr}\left( \hat{\bm\sigma} \dot R^{J\beta}_{\rm n}\left(\frac{\partial {\bm A}^{I\alpha}}{\partial R^{J\beta}_{\rm n}}+ \frac i \hbar  \left[ {\bm A}^{I\alpha},{\bm A}^{J\beta} \right]\right)\right) 
    + \sum_{IJ\beta jk} \hat{\sigma}_{kj}\dot{R}^{J\beta}_{\rm n}\Big[\xi^{J\beta,I\alpha}_{k}\delta_{jk}  -\frac{i}{\hbar}  (\xi^{I\alpha}_{k}-\xi^{I\alpha}_{j})  A^{J\beta}_{jk} \Big] \label{eqn:Ptot-dot-adiab_incomplete1}\\
  &= & \sum_{IJ\beta} \mbox{Tr}\left( \hat{\bm\sigma} \dot R^{J\beta}_{\rm n}\left(-\frac{\partial  {\bm A}^{J\beta}}{\partial R^{I\alpha}_{\rm n}}+\frac{\partial  {\bm A}^{I\alpha}}{\partial R^{J\beta}_{\rm n}}+ \frac i \hbar  \left[  {\bm A}^{I\alpha}, {\bm A}^{J\beta} \right]\right)\right) \label{eqn:Ptot-dot-adiab_incomplete2} \\
 & = & - \sum_{IJ\beta} \mbox{Tr}\left( \hat{\bm\sigma} \dot R^{J\beta}_{\rm n} \bm{\Omega}^{I\alpha J\beta} \right) \label{eqn:Ptot-dot-adiab_incomplete3}
\end{eqnarray}
\end{widetext}
Note that to go from Eq. \ref{eqn:Ptot-dot-adiab_incomplete1} to Eq. \ref{eqn:Ptot-dot-adiab_incomplete2}, we used the following relation, which is proven in Sec. S1 of the SI,
\begin{equation}
    \label{eq:vanish}
    \sum_{I} \frac{ \partial A^{J\beta}_{jk}}{\partial R^{I\alpha}_{\rm n}} =-\sum_{I}\xi^{J\beta,I\alpha}_{k}\delta_{jk}  + \frac{i}{\hbar}\sum_{I}(\xi^{I\alpha}_{k}-\xi^{I\alpha}_{j})A^{J\beta}_{jk}
\end{equation} 

The above analysis leads us to consider the famous
 non-Abelian Berry curvature $\bm{\Omega}^{I\alpha J\beta}$,\cite{Wilczek1984,mead:1992:rmp} which is defined as
\begin{eqnarray}
\label{eqn:non_ab_BC}
    \bm{\Omega}^{I\alpha J\beta} =    \frac{\partial  {{\bm A}^{J\beta}}}{\partial R^{I\alpha}_{\rm n}} - 
     \frac{\partial  {{\bm A}^{I\alpha}}}{\partial R^{J\beta}_{\rm n}} -
     \frac{i}{\hbar} \left[ {{\bm A}^{I \alpha}}, {{\bm A}^{ J \beta}}
    \right]
\end{eqnarray}
As is well-known, the non-Abelian Berry curvature vanishes in the limit of a complete basis, as one can readily demonstrate by inserting a resolution of identity: $\sum_{l}\ket{\psi_{l}}\bra{\psi_l}$, 
\begin{eqnarray}
 \frac{\partial  {A^{J\beta}_{jk}}}{\partial R^{I\alpha}_{\rm n}} - 
     \frac{\partial  {A^{I\alpha}_{jk}}}{\partial R^{J\beta}_{\rm n}} = \frac{i}{\hbar}
 \sum_{l}    \left( A_{jl}^{I\alpha}A_{lk}^{J\beta} -  A_{jl}^{J\beta}A_{lk}^{I\alpha} \right)
 \label{eqn:A_communtator_complete}
\end{eqnarray}
To repeat, the total linear momentum is conserved when we perform the calculation with a complete electronic basis -- and in an adiabatic representation, this conservation becomes clear because the non-Abelian Berry curvature vanishes.

\subsubsection{Angular Momentum Conservation}
Next, let us demonstrate the same conclusion for angular momentum conservation. We 
must evaluate all of the terms in Eq. \ref{eqn:Jtot_dot}.
\begin{itemize}
    \item From the expression for  $\dot {J}^{\alpha}_{\rm n}$  in Eq. \ref{eqn:Jn3} and the expression for $\dot{P}^{I\gamma}_{\rm n}$ in Eq. \ref{eqn:Pn-dot-adiab}, we can write out the first term in Eq. \ref{eqn:Jtot_dot},
    \begin{equation}
\label{eqn:Jn-dot-adiab2}
\begin{split}
        \sum_I \dot {J}_{\rm n}^{I\alpha} &= \sum_{I\beta\gamma} \epsilon_{\alpha\beta\gamma}  
  R^{I\beta}_{\rm n}  \mbox{Tr}\left(\hat{\bm \sigma} \left( -\frac{\partial \hat{\bm V}}{\partial R^{I\gamma}_{\rm n}}  +\frac{i}{\hbar}
 \left[ {\bm A}^{I\gamma},  \hat{\bm V}  \right] \right)  \right)  \\&=  \frac{i}{\hbar} \sum_{I\beta\gamma} \epsilon_{\alpha\beta\gamma}  
  R^{I\beta}_{\rm n}
 \mbox{Tr}\left(\left[ \hat{\bm V}  ,\hat{\bm \sigma}   \right] {\bm A}^{I\gamma}\right) , 
\end{split}
\end{equation}
Above, the first term vanishes because we assume that space is isotropic.
\item Before writing down the expression for the second term  $\mbox{Tr}\left( \hat{\bm \sigma}  \dot {\bm J}_{\rm e}^{\alpha}\right)$ in Eq. \ref{eqn:Jtot_dot}, let us simplify the expression for $ \dot {\bm J}_{\rm e}^{\alpha}$ in Eq. \ref{eqn:AandJ_dot}.
Specifically, we used the following the relation, which is proved in the section S1 of the Supporting Information,
\begin{equation}
\begin{split}
    \label{eq:Jvanish}
     & \sum_{\gamma} \epsilon_{\alpha\eta\gamma} (A_{jk}^{I\gamma} +\xi^{I\gamma}_{k}\delta_{jk} )\\
    &=\sum_{J\beta\gamma} \epsilon_{\alpha\beta\gamma} R^{J\beta}_{\rm n} \Big[-\frac{ \partial A^{I\eta}_{jk}}{\partial R^{J\gamma}_{\rm n}}+\frac{i}{\hbar}(\xi^{J\gamma}_{k} - \xi^{J\gamma}_{j})A^{I\eta}_{jk} -\delta_{jk}\xi^{I\eta,J\gamma}_{k}\Big]
\end{split}
\end{equation}
If we substitute Eq.\ref{eq:Jvanish} into Eq.\ref{eqn:AandJ_dot} and  change dummy index labels,  we recover
\begin{widetext}
    \begin{equation}
\label{eqn:Je_adiabatic_dot}
\begin{split}
        &\mbox{Tr}\left( \hat{\bm \sigma}  \dot {\bm J}_{\rm e}^{\alpha}\right) = \sum_{IJ\eta\beta\gamma jk}\hat{ \sigma}_{kj}\epsilon_{\alpha\eta\gamma}R^{I\eta}_{\rm n} \dot{R}^{J\beta}_{\rm n}\left[\frac{\partial {A}^{I\gamma}_{jk}}{\partial R^{J\beta}_{\rm n}}-\frac{\partial {A}^{J\beta}_{jk}}{\partial R^{I\gamma}_{\rm n}}+ \frac{i}{\hbar}(\xi^{I\gamma}_{k} - \xi^{I\gamma}_{j})A^{J\beta}_{jk}\right]
\end{split}
\end{equation}
\end{widetext}

\item  Using Eq.\ref{eqn:AandJ}, the last term  $\mbox{Tr}\left(\dot {\hat{\bm \sigma}}   {\bm J}_{\rm e}^{\alpha}\right)$ in Eq.\ref{eqn:Jtot_dot} becomes
\begin{equation}
\label{eqn:d_dot_Je_1}
    \mbox{Tr}\left(\dot{\hat{\bm \sigma}}   {\bm J}_{\rm e}^{\alpha}\right)  = \sum_{I\eta\gamma jk} \epsilon_{\alpha\eta\gamma}R^{I\eta}_{\rm n}\dot{\hat{ \sigma}}_{kj}(A^{I\gamma}_{jk}+\xi^{I\gamma}_{k}\delta_{jk})
\end{equation}
We begin by evaluating  plugging Eq. \ref{eqn:d_dot_Pe_2} into  the second term of Eq.\ref{eqn:d_dot_Je_1},
\begin{eqnarray}
\label{eqn:d_dot_Je_2}
   && \sum_{I\eta\gamma jk} \epsilon_{\alpha\eta\gamma}R^{I\eta}_{\rm n}\dot{\hat{ \sigma}}_{kj}\xi^{I\gamma}_{k}\delta_{jk} \nonumber\\&=&  -\frac{i}{\hbar} \sum_{Jjk\beta\gamma\eta }\epsilon_{\alpha\eta\gamma}R^{I\eta}_{\rm n} \dot{R}^{J\beta}_{\rm n} \hat{\sigma}_{kj }(\xi^{I\gamma}_{k}-\xi^{I\gamma}_{j})  A^{J\beta}_{jk}
\end{eqnarray}
If we then plug Eqs. \ref{eqn:d-dot-adiab} and \ref{eqn:d_dot_Je_2} into Eq.\ref{eqn:d_dot_Je_1} and simplify, we find
\begin{widetext}
\begin{equation}
\label{eqn:d_dot_Je}
\begin{split}
   & \mbox{Tr}\left(\dot {\hat{\bm \sigma}}   {\bm J}_{\rm e}^{\alpha}\right)  
 =  \sum_{IJ\eta\gamma\beta} \epsilon_{\alpha\eta\gamma}  
  R^{I\eta}_{\rm n} \Bigg(-\frac{i}{\hbar}
 \mbox{Tr}\left(\left[ \hat{\bm V}  ,\hat{\bm \sigma}   \right] {\bm A}^{I\gamma}\right) +\frac{i}{\hbar} \dot{R}^{J\beta}_{\rm n} \mbox{Tr}\left(\hat{\bm \sigma}[{\bm A}^{I\gamma},{\bm A}^{J\beta}]\right) 
 -  \frac{i}{\hbar}\dot{R}^{J\beta}_{\rm n} \hat{\sigma}_{kj }(\xi^{I\gamma}_{k}-\xi^{I\gamma}_{j})  A^{J\beta}_{jk}\Bigg)
\end{split}
\end{equation}
    
\end{widetext}
\end{itemize}

From Eqs. \ref{eqn:Jn-dot-adiab2}, \ref{eqn:Je_adiabatic_dot}, and \ref{eqn:d_dot_Je}, we can evaluate the time-derivative of angular momentum in Eq. \ref{eqn:Jtot_dot}. Specifically, we see that Eq.\ref{eqn:Jn-dot-adiab2} cancels with the first term in Eq. \ref{eqn:d_dot_Je} and the phase-dependent terms (depending on $\xi$) in Eq.\ref{eqn:Je_adiabatic_dot} and Eq. \ref{eqn:d_dot_Je} cancel as well. The remaining terms  are
\begin{widetext}
   \begin{equation}
\label{eqn:Jtot-dot-adiab_incomplete}
    \begin{split}
    \dot {J}_{\rm tot}^{\alpha}
    =& \sum_{IJ\eta\gamma\beta}
 \epsilon_{\alpha\eta\gamma}  \mbox{Tr}\left(\hat{\bm \sigma} R^{I\eta}_{\rm n}\dot{R}^{J\beta}_{\rm n}\left(\frac{\partial {\bm A}^{I\gamma}}{\partial R^{J\beta}_{\rm n}}-\frac{\partial {\bm A}^{J\beta}}{\partial R^{I\gamma}_{\rm n}} + \frac{i}{\hbar}    \left[{\bm A}^{I\gamma}, {\bm A}^{J\beta} \right]\right)  \right) \\
 =&  - \sum_{IJ\eta\gamma\beta}
 \epsilon_{\alpha\eta\gamma}  \mbox{Tr}\left( \hat{\bm \sigma} R^{I\eta}_{\rm n} \dot R^{J\beta}_{\rm n} {\bm\Omega}^{I\gamma J\beta} \right)
    \end{split}
\end{equation} 
\end{widetext}

As above, the non-Abelian Berry curvature appears and in this case, conservation of angular momentum is implied by the fact that the non-Abelian Berry curvature
vanishes in the limit of a complete set of adiabatic states.

\subsection{Independence of Choice of Gauge}
Before concluding this section, it is crucial to emphasize that the results above do not depend in any way on the gauge $\xi$ in Eqs. \ref{eqn:sum_P}-\ref{eqn:sum_J}.  To the seasoned practitioner of Ehrenfest (or Ehrenfest based) dynamics\cite{miller:2009:jpcareview}, this may not be surprising because Eqs. \ref{eqn:R-dot-adiab}-\ref{eqn:d-dot-adiab}  hold in any basis whatsoever. At the risk of redundancy, for the sake of completeness, let us show this result explicitly by  imagining that we rotate our old set of basis states to a new set of basis states with a unitary matrix $\ket{\psi_{\overline{j}}} =\sum_{k}  \ket{\psi_{k}}\overline{ U}_{k\overline{j}}$. In the new basis, the density matrix, the electronic Hamiltonian, and the Berry connection take the form:
\begin{eqnarray}
    \overline{\bm \sigma} &= & \overline{\bm U}^{\dagger}\hat{\bm \sigma} \overline{\bm U}  \label{eqn:sigma_bar}\\
    \overline{\bm V} &=&\overline{\bm U}^{\dagger} \hat{\bm V}  \overline{\bm U}
    \label{eqn:dVdR_bar} \\
     \overline{\bm A} &=& \overline{\bm U}^{\dagger}{\bm A}  \overline{\bm U} + i\hbar \overline{\bm U}^{\dagger}\nabla_{\rm n}\overline{\bm U}   \label{eqn:A_bar}
\end{eqnarray}
Using Eqs. \ref{eqn:sigma_bar}-\ref{eqn:A_bar} in Eq. \ref{eqn:Pn-dot-adiab}, we can readily show that the equation of motion for the nuclear momentum is unchanged:
\begin{eqnarray}
    \dot{P}^{I\alpha}_{\rm n} 
        &=&-\mbox{Tr}\left( \overline{\bm U}^{\dagger}\hat{\bm \sigma}  \overline{\bm U} \frac{\partial \hat {\bm V} }{\partial R^{I\alpha}_{\rm n}}\overline{\bm U} \right) \nonumber \\
        && + \frac{i}{\hbar}
 \mbox{Tr}\left(\overline{\bm \sigma} \left[\overline{\bm U}^{\dagger} {\bm A}^{I\alpha} \overline{\bm U}  , \overline{\bm V} \right] \right)\label{eqn:Pn-dot-adiab_bar1} \\
 &=& -\mbox{Tr}\left(\overline{\bm \sigma}\frac{\partial \overline{\bm V} }{\partial R^{I\alpha}_{\rm n}}\right)+\frac{i}{\hbar}
 \mbox{Tr}\left(\overline{\bm \sigma} \left[\overline{\bm A}^{I\alpha}  , \overline{\bm V} \right] \right)\label{eqn:Pn-dot-adiab_bar2} 
\end{eqnarray}
Here, we used the relationship:
\begin{equation}
    \nabla_{\rm n}\overline{\bm V}  = \overline{\bm U}^{\dagger} \nabla_{\rm n}\hat{\bm V}\overline{\bm U} - 
\left[\overline{\bm U}^{\dagger}\nabla_{\rm n}\overline{\bm U}  , \overline{\bm V} \right] 
\end{equation}


Next, we can show that the equation of motion for the electronic density matrix propagation is also independent of basis. By definition,
\begin{eqnarray}
\dot{\overline{\bm \sigma}} &=&  \overline{\bm U}^{\dagger}\dot{\hat{\bm \sigma}}\overline{\bm U} + \dot{\bm R}_{\rm n}\left(\nabla_{\rm n}\overline{\bm U}^{\dagger}{\hat{\bm \sigma}}\overline{\bm U} +  \overline{\bm U}^{\dagger}{\hat{\bm \sigma}}\nabla_{\rm n}{\overline{\bm U}}\right) \\
&=&\overline{\bm U}^{\dagger}\dot{\hat{\bm \sigma}}\overline{\bm U} - \left[ \sum_{I}
\frac{{\bm P_{\rm n}^{I}}  }{M_{I}}\cdot \overline{\bm U}^{\dagger}\nabla_{\rm I}\overline{\bm U} , \overline{\bm \sigma} \right] \label{eqn:dbar_dot2}
\end{eqnarray}
Now, using Eqs. \ref{eqn:sigma_bar}-\ref{eqn:A_bar} together with Eq. \ref{eqn:d-dot-adiab}, it follows that:
\begin{eqnarray}
 \overline{\bm U}^{\dagger}\dot{\hat{\bm \sigma}}\overline{\bm U}  &=&-\frac{i}{\hbar} \left[ \overline{\bm V} - \sum_{I}
\frac{{\bm P_{\rm n}^{I}}  }{M_{I}}\cdot  \overline{\bm U}^{\dagger}{\bm A}^{I} \overline{\bm U}, \overline{\bm \sigma} \right]   \\
&=& -\frac{i}{\hbar} \left[ \overline{\bm V} - \sum_{I}
\frac{{\bm P_{\rm n}^{I}}  }{M_{I}}\cdot  \overline{\bm A}^{I}, \overline{\bm \sigma} \right] \nonumber \\
&& + \left[ \sum_{I}
\frac{{\bm P_{\rm n}^{I}}  }{M_{I}}\cdot \overline{\bm U}^{\dagger}\nabla_{\rm I}\overline{\bm U} , \overline{\bm \sigma} \right] 
\label{eqn:dbar_dot1}
\end{eqnarray}
Thus, if we plug Eq. \ref{eqn:dbar_dot1} into Eq. \ref{eqn:dbar_dot2}, we find the desired result:
\begin{equation}
    \dot{\overline{\bm \sigma}} = -\frac{i}{\hbar} \left[ \overline{\bm V} - \sum_{I}
\frac{{\bm P_{\rm n}^{I}}  }{M_{I}}\cdot  \overline{\bm A}^{I}, \overline{\bm \sigma} \right]  \label{eqn:dbar_dot3}
\end{equation}


\section{A Realistic Window With a Truncated Number of Adiabatic Basis Functions}\label{sec:adiab}
In the previous section, we showed that both linear and angular momentum are conserved for the Ehrenfest equations of motion (Eqs. \ref{eqn:R-dot-diab}-\ref{eqn: sigma-dot-diab}) postulated in a complete electronic Hilbert space. In practice, however, a strictly diabatic basis\cite{McLachlan1961,Mead1982,Jasper2004} and 
a complete set of adiabatic states is generally not available; one almost always works in a truncated basis of adiabatic electronic  states. In such a case, 
when studying chemical systems using the formalism above, one might suppose that the linear and angular momentum will not be conserved according to Eqs. ~\ref{eqn:Ptot-dot-adiab_incomplete3} and \ref{eqn:Jtot-dot-adiab_incomplete},  
and thus one might inevitably question the accuracy of such dynamics.

Now, in order to conserve momentum, it is fairly straightforward to guess a solution. Namely, the culprit that has appeared above is the non-Abelian Berry curvature and given Eq. \ref{eqn:Ptot-dot-adiab_incomplete3}, it is fairly straightforward to guess that we ought to damp the nuclear equation of motion by the non-Abelian Berry curvature:
\begin{align}
\label{eqn:guess}
    \ddot{R}^{\alpha}_{\rm n} \rightarrow \ddot{R}^{\alpha}_{\rm n} + \sum_{IJ\beta} \mbox{Tr}\left( \hat{\bm\sigma} \dot R^{J\beta}_{\rm n} \bm{\Omega}^{I\alpha J\beta} \right)
\end{align}
or some variation thereof.  Indeed, such equations have been derived by a  Lagrangian formulation \cite{Amano2005} and a path integral formulation\cite{Krishna2007}. The interested reader can find a proper derivation therein, but for our purposes,  the correct Ehrenfest equations of motion can be heuristically derived from the following effective (non-linear) Hamiltonian $\hat{H}$ in the adiabatic representation,
\begin{equation}
\begin{split}
    \hat{\bm H} =& \sum_{I}\frac{{\bm P^{I}_{\rm n}}^2}{2M_{I}}
-\sum_{I}\frac{{\bm P^{I}_{\rm n}} \cdot {\bm A^{I}}} {M_{I}}
+ \sum_{I}\mbox{Tr}\left( \hat{\bm \sigma}{\bm A^{I}}\right) \cdot \frac{{\bm A^{I}}}{M_{I}}  \\
&-\sum_{I}\frac 1 {2M_{I}} [\mbox{Tr}\left( \hat{\bm \sigma} \bm A^{I}\right)]^2 + \hat{\bm V} 
\end{split}
\label{eqn:H_adiab}
\end{equation}
The corresponding expectation value of the energy $E$ is 
\begin{equation}
\label{eqn:E_adiab}
    E =  
\sum_{I}\frac{{\bm P^{I}_{\rm n}}^2}{2M_{I}}+\mbox{Tr}\left[\hat{\bm \sigma}
\left(\hat{\bm V} -\sum_{I}\frac{{\bm P^{I}_{\rm n}} \cdot {\bm A^{I}}}{M_{I}}\right) \right] +
\sum_{I}\frac 1 {2M_{I}} [\mbox{Tr}\left( \hat{\bm \sigma} \bm A^{I}\right)]^2  
\end{equation}

The equations of motion for $(\bm R_{\rm n}, \bm P_{\rm n})$ are:
\begin{eqnarray}
\label{eqn:eom-adiabatic1}
\dot{\hat{\bm \sigma}} &=& -\frac{i}{\hbar} \left[ \hat{\bm V} -\sum_{I}\frac{{\bm P^{I}_{\rm n}} \cdot {\bm A^{I}}}{M_{I}}
+ \sum_{I}\mbox{Tr}\left(
\hat{\bm \sigma} {\bm A^{I}}
\right) \cdot 
\frac{\bm A^{I}}{M_{I}}, \hat{\bm \sigma} \right]  \\
\label{eq:adrdot}
\dot{R}^{I\alpha}_{\rm n}  &=& \frac{\partial E}{\partial P_{\rm n}^{I\alpha}} = \frac{P_{\rm n}^{I\alpha}}{M_{I}}  - \mbox{Tr}\left( \hat{\bm \sigma}\frac{{\bm A}^{I\alpha}}{M_{I}}  \right) \\
 \dot{P}^{I\alpha}_{\rm n} &=& -\frac{\partial E}{\partial R^{I\alpha}_{\rm n}}\label{eqn:Pn_dot_adiabatic2} \\
 &=&  -\mbox{Tr}\Big(\hat{\bm \sigma}\Big( \frac{\partial \hat{\bm V}}{\partial R^{I\alpha}_{\rm n}}
 -\sum_{J\beta} \frac{[P^{J\beta}_{\rm n}+\mbox{Tr}\left( \hat{\bm\sigma}{\bm A}^{J\beta}\right)]}{M_{J}} \frac{\partial {{\bm A}^{J\beta}}}{\partial R^{I\alpha}_{\rm n}}
 \Big) \Big) \nonumber
\end{eqnarray}
  According to Eq. \ref{eq:adrdot}, we find a difference between the  kinetic and canonical momentum. If we now change variables from the canonical to the  kinetic momentum,\cite{Cotton2017}
\begin{eqnarray}
\label{eqn:pi}
    \pi^{I\alpha}_{\rm n} = P^{I\alpha}_{\rm n} - \mbox{Tr}\left( \hat{\bm\sigma} {\bm A}^{I\alpha} \right),
\end{eqnarray}
we can rewrite the equations of motion in terms of $(\bm R_{\rm n}, \bm \pi_{\rm n})$:
\begin{eqnarray}
\dot{\hat{\bm \sigma}} &=& \frac{-i}{\hbar} \left[ \hat{\bm V} - \sum_{I\alpha}\frac{\bm \pi^{I\alpha}_{\rm n}}{M_{I}} \cdot  \bm A^{I\alpha}, \hat{\bm \sigma} \right]  \label{eqn:d_adia2} \\
\dot{R}^{I\alpha}_{\rm n} &=& \frac{\pi^{I\alpha}_{\rm n}}{M_{I}}\label{eqn:r_adia2}\\
 \dot{\pi}^{I\alpha}_{\rm n} &=& 
 \dot{P}^{I\alpha}_{\rm n} - \mbox{Tr}\left( \hat{\bm \sigma} \dot{{\bm A}}^{I\alpha} \right)
 - \mbox{Tr}\left( \dot{\hat{\bm\sigma}} {\bm A}^{I\alpha} \right)
\label{eqn:pi_dot} \\
 &= &  -\mbox{Tr}\left(\hat{\bm \sigma} \frac{\partial \hat{\bm V}}{\partial R^{I\alpha}_{\rm n}}
   \right) 
   + \sum_{J\beta}\frac{\pi^{J\beta}_{\rm n}}{M_{J}} \mbox{Tr}\left(\hat{\bm \sigma}
 \bm{\Omega}^{I\alpha J\beta} 
 \right) \nonumber \\
 && +  \frac{i}{\hbar} \mbox{Tr} \left( 
  \left[ \hat{\bm V}, \hat{\bm \sigma} \right] {\bm A}^{I\alpha} 
 \right) \label{eqn:pi_adia2}
\end{eqnarray}
These are the equations of motion derived properly in Refs. \citenum{Amano2005,Krishna2007}.
 Compared against the standard adiabatic Ehrenfest equations in Eqs. \ref{eqn:R-dot-adiab}-\ref{eqn:d-dot-adiab}, the equation of motion for the kinetic momentum $\pi^{I\alpha}_{\rm n}$ takes on an additional term that depends on the non-Abelian Berry curvature (in analogy to what was guessed in Eq. \ref{eqn:guess}). Of course, in the limit of a complete set of states, the non-Abelian Berry curvature goes to zero and Eqs. \ref{eqn:d_adia2}-\ref{eqn:pi_adia2} reduce to the standard adiabatic Ehrenfest equations in Eqs. \ref{eqn:R-dot-adiab}-\ref{eqn:d-dot-adiab}. For the sake of completeness (and at the slight risk of redundancy), let us now demonstrate that Eqs. \ref{eqn:d_adia2}-\ref{eqn:pi_adia2} formally obey linear and angular momentum conservation.

\subsection{Linear Momentum Conservation}
We first examine the linear  momentum conservation. Based on the expression for  $\dot{\pi}_{\rm n}^{I\alpha}$ in Eq.\ref{eqn:pi_dot}, 
\begin{eqnarray}
\dot {\bm P}_{\rm tot}^{\alpha} &=&  \sum_{I} \dot{\pi}_{\rm n}^{I\alpha} + \mbox{Tr}\left( \dot{\hat{\bm\sigma}}  {{\bm P}_{\rm e}^{\alpha}} \right)  + \mbox{Tr}\left(  {\hat{\bm\sigma}}  
    \dot {{\bm P}}_{\rm e}^{\alpha} \right) \\
    &=& \sum_{I} \dot{P}_{\rm n}^{I\alpha} + \mbox{Tr}\Big[ \dot{\hat{\bm\sigma}} ( {{\bm P}_{\rm e}^{\alpha} - \sum_{I}{\bm A}^{I\alpha}} )\Big]  \nonumber \\
    && + \mbox{Tr}\Big[   {\hat{\bm\sigma}}  
    ( \dot{\bm P}_{\rm e}^{\alpha} - \sum_{I}{\dot{\bm A}^{I\alpha}} )\Big]  \label{eqn:Ptot_dot2}
\end{eqnarray}
Now evaluate each term in Eq. \ref{eqn:Ptot_dot2}.
\begin{itemize}
    \item  From Eq.\ref{eqn:Pn_dot_adiabatic2}, the first term in Eq. \ref{eqn:Ptot_dot2} is 
    \begin{equation}
    \begin{split}
    \label{eqn:Pn_dot_adiabatic3}
    \sum_{I}\dot  P^{I\alpha}_{\rm n} &= -\sum_{I} \mbox{Tr}\left(\hat{\bm \sigma} \frac{\partial \hat{\bm V}}{\partial R^{I\alpha}_{\rm n}}
   \right) 
   + \sum_{IJ\beta}\frac{\pi^{J\beta}_{\rm n}}{M_{J}} \mbox{Tr}\left(\hat{\bm \sigma}
 \frac{\partial \bm{A}^{J\beta} }{\partial R^{I\alpha}_{\rm n} }    \right) 
 \\&=\sum_{IJ\beta} \dot{R}^{J\beta}_{n} \mbox{Tr}\left(\hat{\bm \sigma}
 \frac{\partial \bm{A}^{J\beta} }{\partial R^{I\alpha}_{\rm n} } \right)     
    \end{split}
\end{equation}
for a translationally invariant potential energy surface.
\item Rearranging Eq. \ref{eqn:AandP} and using Eq. \ref{eqn:d_dot_Pe_2}, the second term in Eq. \ref{eqn:Ptot_dot2} is
\begin{equation}
    \begin{split}
    \label{eqn:Pe_A_diff}
    \mbox{Tr}\Big[ \dot{\hat{\bm\sigma}} ( {{\bm P}_{\rm e}^{\alpha} - \sum_{I}{\bm A}^{I\alpha}} )\Big]& = \sum_{Ijk}\dot{\hat{\sigma}}_{kj} \xi^{I\alpha}_{k}\delta_{jk} \\
    & =  -\frac{i}{\hbar}\sum_{IJ\beta jk}\hat{\sigma}_{kj}\dot{R}^{J\beta}_{\rm n}(\xi^{I\alpha}_{k}-\xi^{I\alpha}_{j})A_{jk}^{J\beta}
        \end{split}
\end{equation}
\item Rearranging Eq. \ref{eqn:AandP_dot}, the third term in Eq. \ref{eqn:Ptot_dot2} is 
\begin{equation}
    \begin{split}
     \label{eqn:Pe_A_diff_dot}
    \mbox{Tr}\Big[   {\hat{\bm\sigma}}  
    ( \dot{\bm P}_{\rm e}^{\alpha} - \sum_{I}{\dot{\bm A}^{I\alpha}} ) \Big] &= \sum_{Ijk}\hat{\sigma}_{kj} \dot{\xi}^{I\alpha}_{k} \delta_{jk} \\&=\sum_{IJ\beta jk}\hat{\sigma}_{kj}\dot{R}^{J\beta}_{\rm n}\xi^{J\beta,I\alpha}_{k}\delta_{jk} 
        \end{split}
\end{equation}
\end{itemize}
If we add   Eq. \ref{eqn:Pn_dot_adiabatic3}, Eq. \ref{eqn:Pe_A_diff}, and Eq. \ref{eqn:Pe_A_diff_dot} together, the time-dependence of the total linear momentum in Eq. \ref{eqn:Ptot_dot2} becomes
\begin{widetext}
\begin{eqnarray}
         \sum_{I} \dot{P}_{\rm tot}^{I\alpha}  
     =  \sum_{IJ\beta} \dot{R}^{J\beta}_{n} \mbox{Tr}\left(\hat{\bm \sigma}
 \frac{\partial \bm{A}^{J\beta} }{\partial R^{I\alpha}_{\rm n} } \right) +\sum_{IJ\beta jk}\hat{\sigma}_{kj}\dot{R}^{J\beta}_{\rm n}\Big(\xi^{J\beta,I\alpha}_{k}\delta_{jk}  -\frac{i}{\hbar} (\xi^{I\alpha}_{k}-\xi^{I\alpha}_{j})A_{jk}^{J\beta}\Big)\label{eqn:Pn_dot_tot_simp2} 
 =0 \label{eqn:Pn_dot_adiab_incomplete}
\end{eqnarray}
\end{widetext}
Here we have used the relation in Eq.\ref{eq:vanish}. In the end,  the proposed equations of motion (Eqs. \ref{eqn:d_adia2}-\ref{eqn:pi_adia2}) strictly conserve linear momentum within a truncated space of adiabatic states.

\subsection{Angular Momentum Conservation}
To investigate angular momentum conservation, we must be very careful now to use Eq. \ref{eqn:Jn1} instead of Eq. \ref{eqn:Jn2} for the nuclear angular momentum, as the two definitions are no longer equivalent. 
The derivative of the total angular momentum is now:
\begin{eqnarray}
\dot {\bm J}_{\rm tot}^{\alpha} &=&  \sum_{\eta\gamma}\epsilon_{\alpha\eta\gamma}  R^{I\eta}_{\rm n} \dot  \pi^{I\gamma}_{\rm n}  +   \mbox{Tr}\left( \hat{\bm\sigma}  \dot {\bm J}_{\rm e}^{\alpha}  + \dot {\hat{\bm \sigma}}   {\bm J}_{\rm e}^{\alpha} \right)\\
&=&  \sum_{I\eta\gamma}\epsilon_{\alpha\eta\gamma}  R^{I\eta}_{\rm n} \dot  P^{I\gamma}_{\rm n} + \mbox{Tr}\left[\hat{\bm\sigma}  \left( \dot {\bm J}_{\rm e}^{\alpha} -\sum_{I\eta\gamma}\epsilon_{\alpha\eta\gamma}  R^{I\eta}_{\rm n}\dot{{\bm A}}^{I\gamma} \right)\right] \nonumber\\
&&+
\mbox{Tr}\left[\dot{\hat{\bm\sigma}}   \left({\bm J}_{\rm e}^{\alpha} -\sum_{I\eta\gamma}\epsilon_{\alpha\eta\gamma}  R^{I\eta}_{\rm n}{{\bm A}}^{I\gamma} \right)\right]
\label{eqn:Jtot_dot2}
\end{eqnarray}
As above, we must evaluate each term in Eq. \ref{eqn:Jtot_dot2}.
\begin{itemize}
    \item  From Eq.\ref{eqn:Pn_dot_adiabatic2}, the first term in Eq. \ref{eqn:Jtot_dot2} can be written as 
    \begin{equation}
    \label{eqn:Jn_dot_adiabatic3}
 \sum_{I\eta\gamma}\epsilon_{\alpha\eta\gamma}  R^{I\eta}_{\rm n} \dot  P^{I\gamma}_{\rm n}  = \sum_{IJjk\eta\beta\gamma}\epsilon_{\alpha\eta\gamma}R^{I\eta}_{\rm n} \dot{R}^{J\beta}_{\rm n}\hat{\sigma}_{kj}\frac{\partial {A}^{J\beta}_{jk}}{\partial R^{I\gamma}_{\rm n}}
\end{equation}
Here we recognize $ - \sum_{I\eta\gamma} \epsilon_{\alpha\eta\gamma}   \mbox{Tr}\left(\hat{\sigma}  R^{I\eta}_{\rm n}\frac{\partial \hat{V}}{\partial R^{I\gamma}_{\rm n}} \right)  = 0 $ due to the isotropy of space.
    \item Rearranging Eq. \ref{eqn:Je_adiabatic_dot}, we recover for the second term in Eq. \ref{eqn:Jtot_dot2}
\begin{equation}
     \begin{split}
\label{eqn:Je_dot_minus_A_dot}
    & \mbox{Tr}\left[\hat{\bm\sigma}  \left( \dot {\bm J}_{\rm e}^{\alpha} -\sum_{I\eta\gamma}\epsilon_{\alpha\eta\gamma}  R^{I\eta}_{\rm n}\dot{{\bm A}}^{I\gamma} \right)\right]\\ =& \sum_{IJjk\eta\beta\gamma}\epsilon_{\alpha\eta\gamma}R^{I\eta}_{\rm n} \dot{R}^{J\beta}_{\rm n}\hat{\sigma}_{kj}\left[-\frac{\partial {A}^{J\beta}_{jk}}{\partial R^{I\gamma}_{\rm n}}+ \frac{i}{\hbar} (\xi^{I\gamma}_{k} - \xi^{I\gamma}_{j})A^{J\beta}_{jk}\right]
     \end{split}
\end{equation}
\item According to Eqs. \ref{eqn:d_dot_Je_1}-\ref{eqn:d_dot_Je_2}, the third term in Eq. \ref{eqn:Jtot_dot2} becomes
\begin{equation}
\begin{split}
\label{eqn:Je_minus_A}    
&\mbox{Tr}\left[\dot{\hat{\bm\sigma}}   \left({\bm J}_{\rm e}^{\alpha} -\sum_{I\eta\gamma}\epsilon_{\alpha\eta\gamma}  R^{I\eta}_{\rm n}{{\bm A}}^{I\gamma} \right)\right]\\
=& -\frac{i}{\hbar}\sum_{IJ\eta\gamma\beta} \epsilon_{\alpha\eta\gamma}  
  R^{I\eta}_{\rm n}    \dot{R}^{J\beta}_{\rm n} \hat{\sigma}_{kj }(\xi^{I\gamma}_{k}-\xi^{I\gamma}_{j})  A^{J\beta}_{jk}
  \end{split}
\end{equation}
\end{itemize}
Comparing Eqs. \ref{eqn:Jn_dot_adiabatic3}-\ref{eqn:Je_minus_A}, we find that when adding them together, the first term of Eq. \ref{eqn:Je_dot_minus_A_dot} cancels with Eq. \ref{eqn:Jn_dot_adiabatic3}. The second term of Eq.\ref{eqn:Je_dot_minus_A_dot} cancels with Eq. \ref{eqn:Je_minus_A}, and hence  $\sum_{I} \dot{J}_{\rm tot}^{I\alpha}  = 0 $.
In the end, with translation symmetry and isotropy of space, propagating the equations of motion in Eqs. \ref{eqn:d_adia2}-\ref{eqn:pi_adia2} conserves both linear and angular momentum -- even for a truncated set of states. 

\subsection{Choice of Gauge and Basis}
As we found when running Ehrenfest dynamics with a complete set of basis states, the result above holds for any choice of gauge in Eqs.\ref{eqn:sum_P}-\ref{eqn:sum_J}; and, more generally, if one considers Eqs. \ref{eqn:d_adia2}-\ref{eqn:pi_adia2}, one finds that these equations are completely unchanged if one rotates the adiabatic states into some other basis set (just as was found for Eqs. \ref{eqn:R-dot-adiab}-\ref{eqn:d-dot-adiab}). The same proof is appropriate, noting only that when we change basis, the non-Abelian Berry curvature has the remarkable property (as shown in section S2 in the SI) that 

\begin{align}
\label{eq:nab_U}
    \overline{\bm {\Omega}}^{I\alpha J\beta} =  \overline{\bm U}^{\dagger}{\bm {\Omega}}^{I\alpha J\beta} \overline{\bm U}
\end{align} 

The Ehrenfest theory above depends only on the window of electronic states chosen (but not on the choice of basis states within that window).

\section{A Different Ehrenfest Approximation}
Before studying several applications of the theory above, it is worth emphasizing that the Eqs.  \ref{eqn:d_adia2}-\ref{eqn:pi_adia2} are not the only possible Ehrenfest approximations. In fact, in the SI, Eqs. S24-S28, we study a different flavor of Ehrenfest approximations \cite{Takatsuka2007,Yonehara2008,Takatsuka2017} where the electronic Hamiltonian and the semiclassical energies in Eqs. \ref{eqn:H_adiab}-\ref{eqn:E_adiab}  are  replaced with
\begin{eqnarray}
  \label{eqn:H_fadiab}
\hat{\bm H} &=& \sum_{I}\frac{{\bm P^{I}_{\rm n}}^2}{2M_I}
-\sum_{I}\frac{{\bm P^{I}_{\rm n}} \cdot {\bm A^{I}}} {M_{I}}
+  \sum_{I}\frac {\bm A^{I^2}} {2M_{I}} + \hat{\bm V} \\
    E &=& \sum_{I}\frac{{ \bm P}_{\rm n}^{I^2}}{2M_{I}}
    +\mbox{Tr}\left[\hat{\bm \sigma} \left(\sum_{I}\frac{{-2\bm P_{\rm n}^{I}} \cdot  {{\bm A}^{I}} + \bm A^{I^2} }{2M_{I}} + \hat{\bm V}\right)\right] \label{eqn:E_fadiab}
\end{eqnarray}

In the SI, we show that the resulting equations still conserve total linear and angular momentum  -- {\em but only with the choice of gauge $\xi = 0$ in Eqs. \ref{eqn:sum_P}-\ref{eqn:sum_J}}. Note that  the resulting dynamics are also {\em not} invariant to changing the adiabatic basis by a unitary transformation.  More discussion regarding the crucial choice of gauge and its implication for {\em ab initio}  on-the-fly dynamics are given below in Sec.\ref{sec:discussion}A.

\section{A Traveling Hydrogen Atom} \label{sec:H}
In Sec.\ref{sec:results} below, we will present an {\em ab initio} calculation exploring momentum conservation numerically. Before presenting such data, however,  it is helpful conceptually to  first treat the simplest, analytical example of Ehrenfest theory: the example of a hydrogen atom traveling at constant velocity. For such a system, $\hat{\bm V}$ and $\bm A$ are constants, and neither the electronic state nor the momentum should  change as a function of time. Thus, as far as the electronic system is concerned, we must find $\dot{\hat{\bm\sigma}} = 0$, which implies that (according to Eq.\ref{eqn:d-dot-adiab}) the electronic wavefunction must be an eigenvector of:
$ \hat{\bm V} - \sum_{\alpha}\dot{R}^{\alpha}_{\rm n} \cdot  \bm A^{\alpha}$. Mathematically, we therefore conclude:
\begin{eqnarray}
\label{eqn:H_commutator}
   \left[ \hat{\bm V}, \hat{\bm \sigma} \right] = \left[  \sum_{\alpha}\dot{R}^{\alpha}_{\rm n} \cdot  \bm A^{\alpha}, \hat{\bm \sigma} \right] 
\end{eqnarray}
so that the electronic population in Eq.\ref{eqn:d-dot-adiab} does not change with the translation of the hydrogen atom. 

Let us now examine how momentum changes with time according to Eq. \ref{eqn:Pn-dot-adiab}. The first term $-\mbox{Tr}\left(\hat{\sigma} \frac{\partial \hat{\bm V}}{\partial R^{\alpha}_{\rm n}}\right)$ in Eq. \ref{eqn:Pn-dot-adiab} is zero due to translation-invariant potential. For the remaining term  in Eq. \ref{eqn:Pn-dot-adiab}, $ \frac{i}{\hbar} \mbox{Tr} \left( 
  \left[ \hat{\bm V}, \hat{\bm \sigma} \right] {\bm A}^{\alpha} 
 \right) $, we plug in Eq. \ref{eqn:H_commutator}, and Eq. \ref{eqn:Pn-dot-adiab} then becomes
\begin{eqnarray}
    \dot{\bm P}_{\rm n}^{\alpha} &=& \frac{i}{\hbar} \mbox{Tr} \left( 
  \left[ \hat{\bm V}, \hat{\bm \sigma} \right] {\bm A}^{\alpha} 
 \right)   \\
 & = & \frac{i}{\hbar}  \sum_{\beta} \dot{R}^{\beta}_{\rm n} \mbox{Tr} \left(\hat{\bm \sigma}  \left[ {\bm A}^{\alpha},\bm A^{\beta}  \right] \right) \label{eqn:A_commutator}
\end{eqnarray}
As the derivative couplings between certain eigenstates of the hydrogen atom are non-zero (i.e. $1s$ and $3p_{x}$), Eq.\ref{eqn:A_commutator} is non-zero. Therefore, using the standard Ehrenfest approach in Eqs.\ref{eqn:R-dot-adiab}-\ref{eqn:d-dot-adiab} cannot capture a translating hydrogen atom in a truncated basis: an isolated hydrogen atom changes its momentum during translation. 

Now there are two ways to resolve this issue. $(i)$ One way is to apply  electron-translation factor corrections, \cite{Fatehi2011,Athavale2023}  which effectively allows us to replace the Berry connection by zero, $\bm A \rightarrow 0$ in Eqs. \ref{eqn:Pn-dot-adiab}-\ref{eqn:d-dot-adiab}, so that the wavefunction $\phi(r,t) = \phi_{1s}(r-R(t))$ becomes a stationary state (though admittedly without any electronic momentum). \cite{Nafie1983,Patchkovskii2012}
Alternatively, $(ii)$ a second approach is to  use the effective Ehrenfest equations in Eqs. \ref{eqn:d_adia2}-\ref{eqn:pi_adia2}, according to which the kinetic momentum feels a force from the  non-Abelian Berry curvature (Eq. \ref{eqn:non_ab_BC}), so that the second term of Eq. \ref{eqn:pi_adia2} becomes
\begin{eqnarray}
&&\sum_{\beta}\frac{\pi^{\beta}_{\rm n}}{M} \mbox{Tr}\left(\hat{\bm \sigma}
 \bm{\Omega}^{\alpha \beta} 
 \right)  \nonumber\\
 &=& \sum_{\beta} \frac{\pi^{\beta}_{\rm n}}{M}  \mbox{Tr}\left\{\hat{\bm \sigma} \left( \frac{\partial  {{\bm A}^{\beta}}}{\partial R^{\alpha}_{\rm n}} - 
     \frac{\partial  {{\bm A}^{\alpha}}}{\partial R^{\beta}_{\rm n}}
-\frac{i}{\hbar}  \left[ {\bm A}^{\alpha},\bm A^{\beta}  \right]
 \right) \right\}
\end{eqnarray}
Moreover, for a hydrogen atom, $\bm A$ does not depend on $\bm R_{\rm n}$ so that 
\begin{eqnarray}
\label{eqn:A_zero}
\sum_{\beta}\frac{\pi^{\beta}_{\rm n}}{M} \mbox{Tr}\left(\hat{\bm \sigma}
 \bm{\Omega}^{\alpha \beta} 
 \right) = -\frac{i}{\hbar} \sum_{\beta} \frac{\pi^{\beta}_{\rm n}}{M}  \mbox{Tr}\left\{\hat{\bm \sigma} \left( 
  \left[ {\bm A}^{\alpha},\bm A^{\beta}  \right]
 \right) \right\}
\end{eqnarray}
The first term on the right-hand side of Eq. \ref{eqn:pi_adia2} is still zero, and thus using Eq. \ref{eqn:A_commutator} and Eq. \ref{eqn:A_zero}, the right-hand side of Eq. \ref{eqn:pi_adia2} is entirely zero. In other words, in order for Ehrenfest dynamics to properly capture a traveling hydrogen atom, one requires inserting the non-Abelian curvature in the equation of motion for the momentum. Presumably, this non-Abelian Berry curvature is not needed in the limit of a  complete electronic basis.


\section{Results} \label{sec:results}
\begin{figure}[h]
\includegraphics[width=3.3in]{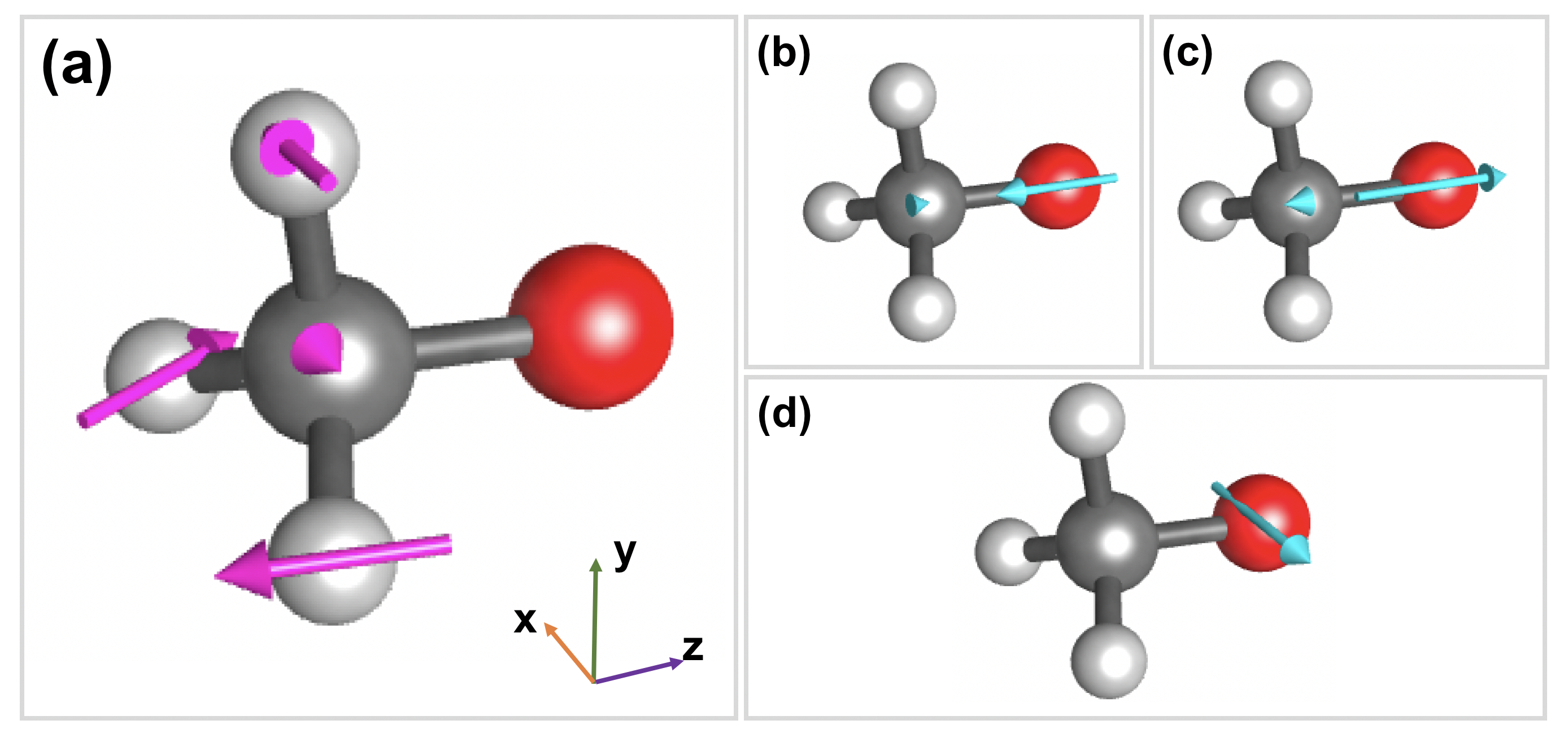}
\caption{\label{fig:geom} (a) [Magenta] The displacements of the first vibrational mode  along which we propagate dynamics.  In (b), we plot [cyan] the atom-based spin magnetic moments for a single GHF+SOC solution,
and in (c) we plot the corresponding moments for its degenerate time-reversed solution.
(d) The weighted spin magnetic moments on each atom using the initial amplitude $\bm  c = (\frac{1}{\sqrt{2}}, \frac{i}{\sqrt{2}})$. }
\end{figure}

To verify and further investigate  angular momentum conservation, we have performed Ehrenfest dynamics with (Eqs. \ref{eqn:d_adia2}-\ref{eqn:pi_adia2})
and without the Berry force (Eqs.\ref{eqn:R-dot-adiab}-\ref{eqn:d-dot-adiab}) for the methoxy radical, a doublet that exhibits a Kramers degeneracy.  The initial geometry was optimized with the hydrogens on the carbon at the unrestricted Hartree-Fock level of theory. We calculated the relevant ground state using generalized Hartree-Fock (GHF) theory -- i.e., we assume a HF ansatz where each orbital can be a linear combination of a spin up and spin down spatial orbtial so that $S_z$ is no longer a good quantum number. We used the 6-31G basis set and we included SOC.\cite{Abegg1975,Tao2023} GHF is the HF equivalent of non-collinear density-functional theory.\cite{Kubler1988} Note that the GHF ansatz converges to one state of the doublet (herefter denoted $\ket{GHF}$; the other state was generated by applying time reversal symmetry operator (herefafter denoted $\ket{TGHF})$. Note also  that because the energies of the Kramers doublet ground states are degenerate, the last term in Eq.\ref{eqn:pi_adia2} is zero.  The initial velocity was set to be the direction corresponding to the lowest vibrational mode  with an initial kinetic energy of 0.005 a.u. ($\approx 1491K$); see Fig.\ref{fig:geom}(a).  The dynamics were performed with a step size of 5 a.u. (0.121 fs).  The initial amplitude was fixed as $\bm  c = (\frac{1}{\sqrt{2}}, \frac{i}{\sqrt{2}})$ in the GHF/TGHF basis, which gives an initial density matrix $\bm  \sigma = \bm  c\bm  c^{T} = \begin{pmatrix} \frac{1}{2},\frac{i}{2}\\-\frac{i}{2},\frac{1}{2}\end{pmatrix}$. The non-Abelian Berry curvature in Eq. \ref{eqn:pi_adia2} was computed by finite-difference. The calculations were performed in a local branch of Q-Chem 6.0.\cite{Epifanovsky2021} 

In Fig.\ref{fig:geom}(b) and (c), using cyan arrows, we plot the spin magnetic  moments on each atom according to a Mulliken-like scheme\cite{Jiménez-Hoyos2014} at time zero for each of the two double states; one doublet state is plotted in (b), and the time reversed state is plotted in (c) (which is of course in the exact opposite direction). In Fig.\ref{fig:geom}(d), we plot the weighted spin magnetic moments on each atom using the initial amplitude, $\bm  c = (\frac{1}{\sqrt{2}}, \frac{i}{\sqrt{2}})$.  In Fig. S1, we plot the change in the amplitudes and the population during dynamics. The Ehrenfest average of the atom-based spin magnetic moments rotate in the xy plane, as also shown in Fig. S1 in the Supporting Information.

\begin{figure*}
\includegraphics[width=6in]{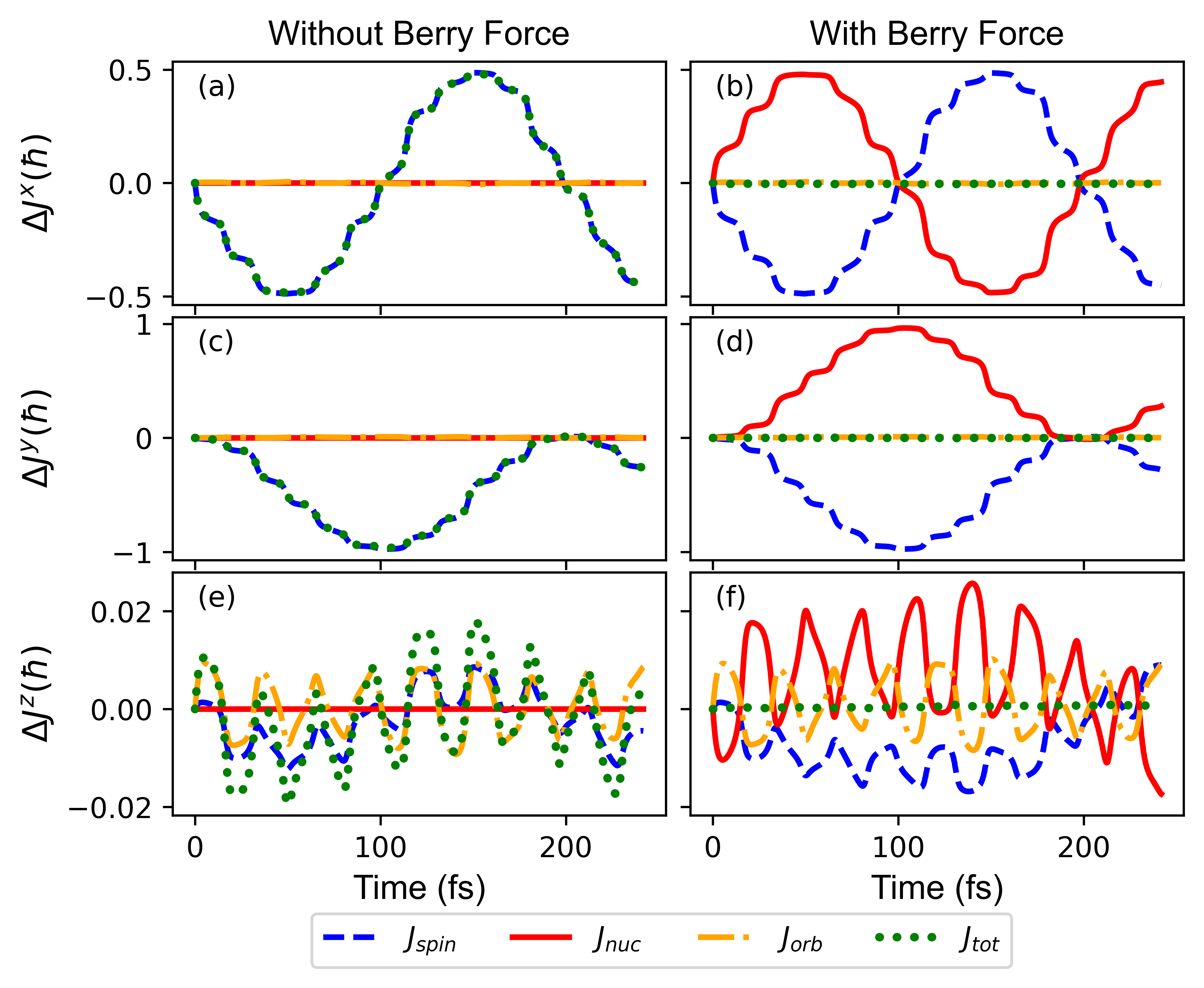}
\caption{\label{fig:angmom} The change in the real-time  angular momentum $\Delta J^\alpha (t) =   J^\alpha (t)-  J^\alpha (0)$ (relative to time zero) according to
an Ehrenfest trajectory simulating methoxy radical moving along its lowest normal mode vibration. The nonzero initial values of components of the angular momentum are listed: $J^{y}_{\rm spin} = 0.4811\hbar$, $J^{y}_{\rm orb} = -0.0054\hbar$, $J^{\rm x}_{\rm nuc} = 0.0075\hbar$, and $J^{\rm z}_{\rm nuc} = -0.0048\hbar$. The three panels (a),(c),(e) on the left exclude the non-Abelian Berry force and  correspond to the Cartesian coordinates $x$, $y$ and $z$ shown in Fig.~\ref{fig:geom} (and the same for Figs.(b),(d) and (f) but now including the Berry force). Without a Berry force, the total angular momentum is not conserved; with a Berry force, the total angular momentum is conserved. In Figs.(b) and (d), when we correctly include the Berry force, we find that the nuclear and spin degrees of freedom transfer angular momentum between each other.   In Figs (f), the nuclear, spin, and electronic orbital degrees of freedom all exchange angular momentum.  Comparing (e) and (f), we observe that the electronic spin changes differently with or without a Berry force; this change arises because adding the Berry force can changes the nuclear motion and yields a reasonably different trajectory. }
\end{figure*} 

In Fig.\ref{fig:angmom}, we plot the change in the angular momentum and linear momentum during the trajectory. 
To begin our discussion, consider first the case where
Berry force is not included (Fig.\ref{fig:angmom}(a)(c)(e))).
When the Berry force is not included, the nuclear angular momentum is  calculated from the canonical momentum $\bm P_{\rm n}$, which is conserved for a translationally invariant surface, and hence there is no change in the nuclear momentum (red solid line). Note that the canonical momentum is conserved in this example because the two basis states are degenerate. In the more general case, with multiple non-degenerate states, the canonical momentum would not be conserved, as shown previously in Ref.\cite{Shu2020}.  In blue, we pot the electronic spin angular momentum; in orange, we plot the electronic orbital angular momentum.  In Fig.\ref{fig:angmom}(a) and (c), the electronic spin angular momentum changes tracks exactly with the total angular momentum (i.e. the electronic orbital angular momentum is effectively constant). In Fig.\ref{fig:angmom}(e), both the electronic orbital and spin angular momentum fluctuate, and the total angular momentum appears more chaotic. 

Next, consider Fig.\ref{fig:angmom}(b)(d)(f)) where the Berry force is included. Here,  we see the nuclear angular momentum (as calculated from the kinetic momentum $\bm \pi_{\rm n}$ in Eq. \ref{eqn:pi_adia2}) changes with time, and the  Berry force captures the angular momentum transfer from the electronic spin/orbital angular momentum to the nuclear angular momentum.  As must be true, the total angular momentum is constant and  conserved.
(In Fig.S2, we also show numerically that the total linear momentum is conserved when a Berry force is included.) 
Of most importance, when comparing Figs. \ref{fig:angmom}(e) and (f), we observe that the electronic spin changes noticeably depending on whether or not a Berry force is included, clearly emphasizing the importance of going beyond standard BO dynamics (and including Berry forces) in the presence of non-trivial spin degrees of freedom.

\section{Discussion}
\label{sec:discussion}
\subsection{Choice of Phase/Gauge $\xi$}
At this point, it is essential for us to discuss our choice of phase.  For the case of a real Hamiltonian, one can choose the Hamiltonian eigenfunctions to be real as well (in a smooth fashion), and thus one can ignore the gauge choice $\xi$ in Eqs. \ref{eqn:sum_P}-\ref{eqn:sum_J}. However, in the case of a complex-valued Hamiltonian, the choice of phase is far more complicated. Obviously, Berry phases can appear (which should not be ignored) if one moves around in a closed loop\cite{Berry1984}. Even more importantly for our semiclassical purposes, the choice of phase will always be somewhat uncontrollable for {\em ab initio} on-the-fly dynamics
because one must pick the phase of the resulting wavefunction at each step with very limited information: one does not have the capacity to make sure that the phases of wavefunctions are matched for similar nuclear configurations and  one cannot easily attach different phase factors for translational, rotational, and internal motion. Thus, at the end of the day, for the most part, the usual approach is simply to align the phases of nuclear wavefunctions at two slightly different geometries (separated by one time step) using parallel transport. Parallel transport does not satisfy $\xi = 0$ in Eqs. \ref{eqn:sum_P}-\ref{eqn:sum_J}. Therefore, when running semiclassical dynamics, one seeks equations of motion that are insensitive to the  gauge and to that end, as a practical matter,  the Ehrenfest equations in Eqs. \ref{eqn:d_adia2}-\ref{eqn:pi_adia2} have a huge advantage over those in Eqs. \ref{eqn:H_fadiab}-\ref{eqn:E_fadiab}.

\begin{figure*}
\includegraphics[width=6in]{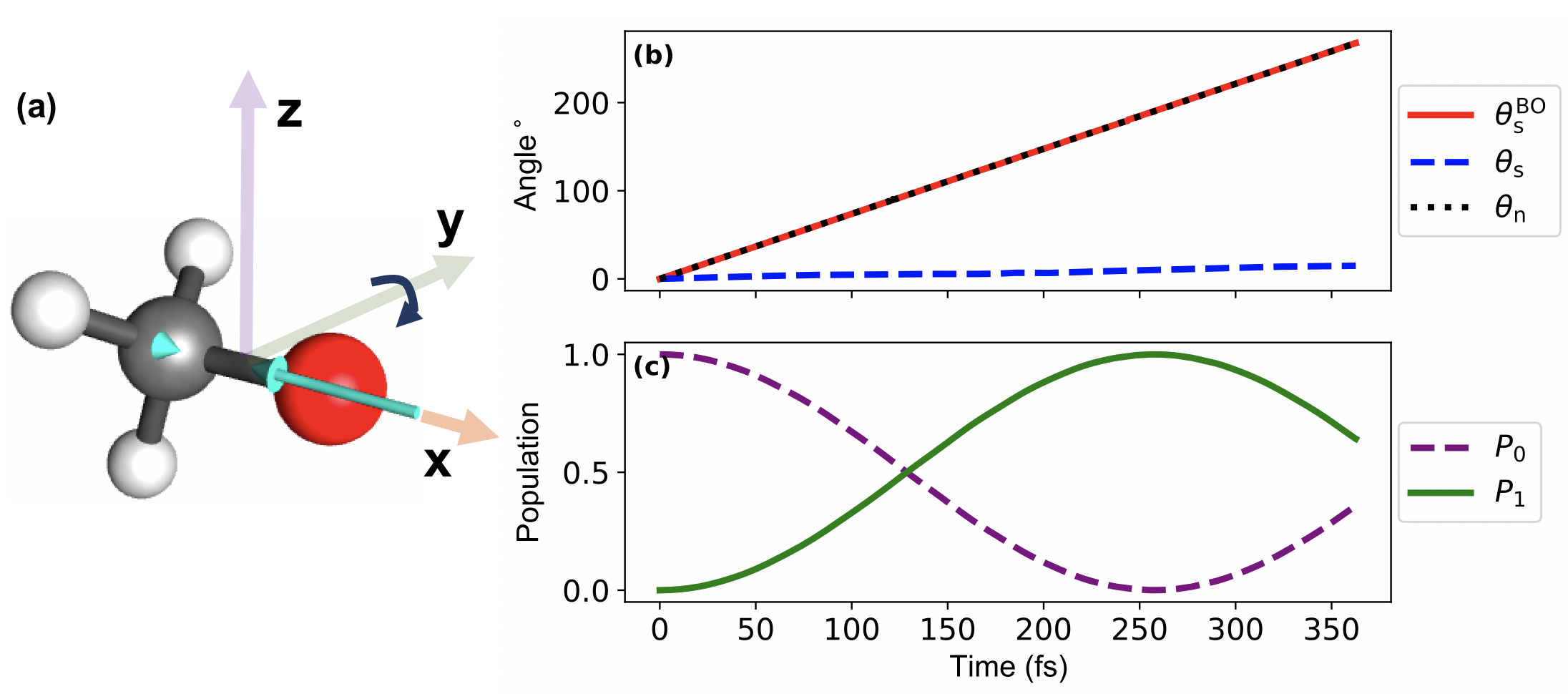}
\caption{\label{fig:rotate}  (a) The orientation of the methoxy radical  molecule, the rotation direction, and atom-based spin magnetic moments (cyan arrows) at the initial step. (b) The rotation angles of the molecule (black dotted line), the total spin vector from the BO dynamics (red solid line), and from the Ehrenfest dynamics (blue dashed lines). (c) The populations for the Kramers doublet states ($P_{0}$ and $P_{1}$) with time from the Ehrenfest dynamics. }
\end{figure*}

\subsection{Ehrenfest vs BO Dyanmics}
Lastly, before concluding, in order to numerically emphasize the need to go beyond the BO approximation when treating spin spin degrees of freedom, we will report one more simulation. Let us orient the methoxy radical molecule with the CO bond aligned along the x axis; within such a frame, a GHF calculation with SOC reveals that all spin magnetic moments point along the x-axis (Fig. \ref{fig:rotate}(a)). Let us  now apply a rotational force around the y axis,  with the initial nuclear angular momentum $J_{\rm nuc} = (0.0, 39.28\hbar, 0.0)$,
and propagate the resulting dynamics with both BO and Ehrenfest. For both sets of dynamics, we include the corresponding 
Berry force (the on-diagonal Berry force for BO dynamics\cite{Tao2023,Bian2023} and the non-Abelian Berry force \cite{Amano2005,Krishna2007} for Ehrenfest dynamics), so that both trajectories conserve the total angular momentum.  For additional trajectory data, and in particular for the time-dependent state populations and an analysis of the spin angular momentum in terms of the relevant  $\ket{GHF}$ and $\ket{TGHF}$ wavefunctions,  see the SI. 
In Fig. \ref{fig:rotate}(b), as a function of time, we plot the rotation angle for the molecule as well as the total expectation values for the spin vectors according to both BO and Ehrenfest dynamics. 
As one might expect, within the BO approximation (as calculated along a continuous GHF+SOC state), the total spin vector $\textbf{S}_{\rm BO}$  rotates with the molecule (red line).
This prediction is of course unphysical, however; the spin direction does not change instantaneously with molecular frame
but rather changes depending on the spin-orbit coupling. This slow change of direction is correctly captured by Ehrenfest dynamics (blue dashed lines). Lastly, this BO failure can be verified in \ref{fig:rotate}(c), where we plot population as a function of time and show, the time the molecule has rotated 180 degrees (244 fs), 
the populated Kramers doublet has effectively switched, 
which represents a complete breakdown of the BO approximation.

\section{Conclusions}\label{sec:conclusion}

In this paper, we have demonstrated that, in order for Ehrenfest dynamics to conserve the linear or angular momentum in a truncated basis, the nuclei must experience a force arising from the non-Abelian curvature $\bm{\Omega}$ (as present in Eqs. \ref{eqn:d_adia2}-\ref{eqn:pi_adia2}). This result is independent of the choice of gauge for the electronic states in Eqs. \ref{eqn:sum_P}-\ref{eqn:sum_J}.  
As examples, we have studied both the traveling hydrogen atom (analytically) as well as the methoxy radical (numerically). Both examples make clear that  the non-Abelian curvature term is needed and, in the case of the methoxy radical, our data also highlights that the evolution of the spin degrees of freedom can be different as a result.   Looking forward, the present results should have immediate impact in a variety of fields where nuclear, electronic and spin motion are all entangled perhaps especially for systems displaying chiral-induced spin selectivity\cite{Das2022,Evers2022}.

\section{Supplementary Material}
In the  supplementary material, we include a proof of Eqs.\ref{eq:vanish} and \ref{eq:Jvanish}, a proof of the gauge covariance of the non-Abelian berry curvature (Eq.\ref{eq:nab_U}), a detailed discussion of the alternative Ehrenfest scheme in Eqs.\ref{eqn:H_fadiab}-\ref{eqn:E_fadiab} vis-a-vis momentum conservation, and more details of the methoxy radical dynamics.

\section{Acknowledgment}
We would like to thank Tanner Culpitt for a careful reading. This work is supported by the National Science Foundation under Grant No. CHE-2102402.

\bibliography{main}

\end{document}


\newpage
\tableofcontents

\newpage
\section{Proof of Eq. 40 and Eq. 44}
In this section we prove the relationships in Eqs. 40 and 44 from the main paper. For convenience, we also write them as Eq.~(\ref{eqn:vanish}) and Eq.~(\ref{eqn:Jvanish}) here, respectively.
\begin{eqnarray}
        \label{eqn:vanish}
    \sum_{I} \frac{ \partial A^{J\beta}_{jk}}{\partial R^{I\alpha}_{\rm n}} &=&-\sum_{I}\xi^{J\beta,I\alpha}_{k}\delta_{jk}  + \frac{i}{\hbar}\sum_{I}(\xi^{I\alpha}_{k}-\xi^{I\alpha}_{j})A^{J\beta}_{jk} \\
        \label{eqn:Jvanish}
      \sum_{\gamma} \epsilon_{\alpha\beta\gamma} (A_{jk}^{I\gamma} + \xi^{I\gamma}_{k}\delta_{jk} )
    &=&\sum_{J\eta\gamma} \epsilon_{\alpha\eta\gamma} R^{J\eta}_{\rm n} \Big[-\frac{ \partial A^{I\beta}_{jk}}{\partial R^{J\gamma}_{\rm n}}+\frac{i}{\hbar}(\xi^{J\gamma}_{k} - \xi^{J\gamma}_{j})A^{I\beta}_{jk} -\delta_{jk}\xi^{I\beta,J\gamma}_{k}\Big]
\end{eqnarray}
We start by proving Eq.~(\ref{eqn:vanish}) by taking the derivative of the Berry connection:
\begin{eqnarray}
        \sum_{I} \frac{\partial A^{J\beta}_{jk}}{\partial R^{I\alpha}_{\rm n}} & =& 
        i\hbar \sum_{I}\bra{\nabla_{I\alpha}\psi_{j}}\ket{\nabla_{J\beta}\psi_{k}}  + i\hbar\bra{\psi_{j}}\nabla_{J\beta}\sum_{I}\nabla_{I\alpha}\ket{\psi_{k}} \\
         & =& i\hbar\sum_{I}\bra{\nabla_{I\alpha}\psi_{j}}\ket{\nabla_{J\beta}\psi_{k}}  - \bra{\psi_{j}}\nabla_{J\beta}\sum_{I}\hat{P}_{\rm n}^{I\alpha}\ket{\psi_{k}} \label{eqn:A_Rdependence1}
\end{eqnarray}
Recall the smooth gauge convention for the adiabatic states, as given in the Eq.~(25) in the main paper, 
\begin{equation}
    \label{eqn:sum_P}
 \Big(\sum_{I}\hat { P}^{I\alpha}_{\rm n} +  \hat { P}^{\alpha}_{\rm e}\Big)\ket{\psi_{k}} = \sum_{I}\xi^{I\alpha}_{k}(\bm R_{\rm n})\ket{\psi_{k}} .
\end{equation}
From this phase convention, the second term in Eq.~(\ref{eqn:A_Rdependence1}) becomes:
\begin{eqnarray}
   && -\sum_{I}\bra{\psi_{j}}\nabla_{J\beta}\hat{P}_{\rm n}^{I\alpha}\ket{\psi_{k}} \nonumber \\&=&
 \bra{\psi_{j}}\nabla_{J\beta}(\hat{P}_{\rm e}^{\alpha}-\sum_{I}\xi^{I\alpha}_{k})\ket{\psi_{k}}  \\
 &=&
 \bra{\psi_{j}}\hat{P}_{\rm e}^{\alpha}\ket{\nabla_{J\beta}\psi_{k}} - \sum_{I}\bra{\psi_{j}}(\nabla_{J\beta}\xi^{I\alpha}_{k})\ket{\psi_{k}} -\sum_{I}\xi^{I\alpha}_{k} \bra{\psi_{j}}\nabla_{J\beta}\ket{\psi_{k}}  \\
        & =&  \bra{(\xi^{\alpha}_{j}- \hat{P}^{\alpha}_{\rm n})\psi_{j}}\ket{\nabla_{J\beta}\psi_{k}} -\sum_{I}\xi^{J\beta,I\alpha}_{k}\delta_{jk}+\frac{i}{\hbar}\sum_{I} \xi^{I\alpha}_{k} A^{J\beta}_{jk}\\
        & =&  -i\hbar \sum_{I}\bra{\nabla_{I\alpha}\psi_{j}}\ket{\nabla_{J\beta}\psi_{k}}-\sum_{I}\xi^{J\beta,I\alpha}_{k}\delta_{jk}+\frac{i}{\hbar}\sum_{I}(\xi^{I\alpha}_{k}-\xi^{I\alpha}_{j})  A^{J\beta}_{jk} \label{eqn:A_Rdependence2}
\end{eqnarray}
If we plug Eq.~(\ref{eqn:A_Rdependence2}) into Eq.~(\ref{eqn:A_Rdependence1}), then we obtain Eq.~(\ref{eqn:vanish}).

Next we prove  Eq.~(\ref{eqn:Jvanish}). Note that Eq.~(\ref{eqn:Jvanish}) can be re-arranged as
\begin{equation}
    \label{eqn:RXA_Rdependence}
    \sum_{J\gamma\eta} \epsilon_{\alpha\eta\gamma} R^{J\eta}_{\rm n}\frac{ \partial A^{I\beta}_{jk}}{\partial R^{J\gamma}_{\rm n}} = \sum_{J\eta\gamma} \epsilon_{\alpha\eta\gamma} R^{J\eta}_{\rm n} \Big[\frac{i}{\hbar}(\xi^{J\gamma}_{k} - \xi^{J\gamma}_{j})A^{I\beta}_{jk} -\delta_{jk}\xi^{I\beta,J\gamma}_{k}\Big] - \sum_{\gamma} \epsilon_{\alpha\beta\gamma} (A_{jk}^{I\gamma} +\xi^{I\gamma}_{k}\delta_{jk} )
\end{equation}
The left-hand-side of Eq.~(\ref{eqn:RXA_Rdependence}) can be expanded
\begin{eqnarray}
    & & \sum_{J\gamma\eta} \epsilon_{\alpha\eta\gamma} R^{J\eta}_{\rm n}\frac{ \partial A^{I\beta}_{jk}}{\partial R^{J\gamma}_{\rm n}}\nonumber\\ 
    &= &  i\hbar \sum_{J\gamma\eta} \epsilon_{\alpha\eta\gamma}\left( \bra{\psi_j}\ket{R^{J\eta}_{\rm n} \nabla_{I\beta}\nabla_{J\gamma} \psi_k} +  \bra{R^{J\eta}_{\rm n}\nabla_{J\gamma} \psi_j}\ket{\nabla_{I\beta}\psi_k} \right)\\\ 
    &=&  i\hbar   \sum_{J\gamma\eta} \epsilon_{\alpha\eta\gamma}\bra{\psi_j} \left( \nabla_{I\beta} R^{J\eta}_{\rm n} + \left[ R^{J\eta}_{\rm n}, \nabla_{I\beta}  \right] \right) \nabla_{J\gamma}  
    \ket{\psi_k} +  \bra{\hat{J}_{\rm n}^{\alpha} \psi_j}\ket{\nabla_{I\beta}\psi_k}\label{eqn:commutator}\\  
    &=&  -\bra{\psi_j}\nabla_{I\beta} \hat J_{\rm n}^{\alpha} \ket{\psi_k} - \sum_{\gamma} \epsilon_{\alpha\beta\gamma} A^{I\gamma}_{jk} +  \bra{\hat{J}_{\rm n}^{\alpha}  \psi_j}\ket{\nabla_{I\beta}\psi_k} 
    \label{eqn:RXA_Rdependence1}
\end{eqnarray}
In moving from Eq.~(\ref{eqn:commutator}) to Eq.~(\ref{eqn:RXA_Rdependence1}), we have used the commutator relationship, $\left[ R^{J\eta}_{\rm n}, \nabla_{I\beta}  \right] = -\delta_{IJ}\delta_{\beta\eta}$. 

Next, recall the gauge convention for the adiabatic states, as given in the Eq.~(26) in the main paper:
\begin{equation}
\label{eqn:sum_J}
 \Big(\sum_{I}\hat { J}^{I\alpha}_{\rm n}+  \hat { J}^{\alpha}_{\rm e}\Big)\ket{\psi_{k}} =  \sum_{I\eta\gamma} \epsilon_{\alpha\eta\gamma} R^{I\eta}_{\rm n}\xi^{I\gamma}_{k}(\bm R_{\rm n})\ket{\psi_{k}} 
\end{equation}
from which the last term in Eq.~(\ref{eqn:RXA_Rdependence1}) becomes:
\begin{eqnarray}
   && \bra{\hat{J}_{\rm n}^{\alpha} \psi_j}\ket{\nabla_{I\beta}\psi_k} \nonumber\\&=&   \bra{\Big(-\hat{J}_{\rm e}^{\alpha} + \sum_{J\eta\gamma}\epsilon_{\alpha\eta\gamma} R^{J\eta}_{\rm n}\xi^{J\gamma}_{j}\Big)\psi_j}\ket{\nabla_{I\beta}\psi_k} \\
    & =& -\bra{\hat{J}_{\rm e}^{\alpha} \psi_j}\ket{\nabla_{I\beta}\psi_k}+\sum_{J\eta\gamma} \epsilon_{\alpha\eta\gamma} R^{J\eta}_{\rm n}  \xi^{J\gamma}_{j}\bra{\psi_j}\ket{\nabla_{I\beta}\psi_k} \\
    & = &-\bra{\psi_j}\ket{\nabla_{I\beta}\hat{J}_{\rm e}^{\alpha} \psi_k} -\frac{i}{\hbar}\sum_{J\eta\gamma} \epsilon_{\alpha\eta\gamma} R^{J\eta}_{\rm n}  \xi^{J\gamma}_{j}A^{I\beta}_{jk} \\    &=&\bra{\psi_j}\ket{\nabla_{I\beta}\Big(\hat{J}_{\rm n}^{\alpha}- \sum_{J\eta\gamma}\epsilon_{\alpha\eta\gamma} R^{J\eta}_{\rm n}\xi^{J\gamma}_{k}\Big)\psi_k} -\frac{i}{\hbar}\sum_{J\eta\gamma} \epsilon_{\alpha\eta\gamma} R^{J\eta}  \xi^{J\gamma}_{j}A^{I\beta}_{jk} \\
    & =& \bra{\psi_j}\nabla_{I\beta} \hat J_{\rm n}^{\alpha} \ket{\psi_k} -\sum_{J\eta\gamma}\epsilon_{\alpha\eta\gamma} \bra{\psi_j}\nabla_{I\beta} R^{J\eta}_{\rm n}\xi^{J\gamma}_{k}\ket{\psi_k}-\frac{i}{\hbar}\sum_{J\eta\gamma} \epsilon_{\alpha\eta\gamma} R^{J\eta}_{\rm n}  \xi^{J\gamma}_{j}A^{I\beta}_{jk} \label{eqn:RXA_Rdependence2}
\end{eqnarray}
At this point, let us write out the middle term in Eq.~(\ref{eqn:RXA_Rdependence2}),
\begin{eqnarray}
    \bra{\psi_j}\nabla_{I\beta} R^{J\eta}_{\rm n}\xi^{J\gamma}_{k}\ket{\psi_k} =   \delta_{IJ}\delta_{\eta\beta}\xi^{J\gamma}_{k} \delta_{jk}+ R^{J\eta}_{\rm n}(\nabla_{I\beta} \xi^{J\gamma}_{k} ) \delta_{jk} + R^{J\eta}_{\rm n}\xi^{J\gamma}_{k} \bra{\psi_j}\ket{\nabla_{I\beta}\psi_k} \label{eqn:RXA_Rdependence3}
\end{eqnarray}
Note that here we evaluated $\nabla_{I\beta}  R^{J\eta}_{\rm n} = \delta_{IJ}\delta_{\eta\beta}$. 

The derivation is almost over. If we plug Eq.~(\ref{eqn:RXA_Rdependence3}) into Eq.~(\ref{eqn:RXA_Rdependence2}),  Eq.~(\ref{eqn:RXA_Rdependence2}) becomes
\begin{equation}
\label{eqn:RXA_Rdependence4}
    \bra{\hat{J}_{\rm n}^{\alpha} \psi_j}\ket{\nabla_{I\beta}\psi_k} = \bra{\psi_j}\nabla_{I\beta} \hat J_{\rm n}^{\alpha} \ket{\psi_k}- \sum_{\beta\gamma}\epsilon_{\alpha\beta\gamma} \xi^{I\gamma}_{j} \delta_{jk} +\sum_{J\eta\gamma} \epsilon_{\alpha\eta\gamma} R^{J\eta}_{\rm n} \Big[\frac{i}{\hbar}(\xi^{J\gamma}_{k} - \xi^{J\gamma}_{j})A^{I\beta}_{jk} - \delta_{jk}\xi^{I\beta,J\gamma}_{k}\Big]
\end{equation}
Lastly, we plug Eq.~(\ref{eqn:RXA_Rdependence4}) back into Eq.~(\ref{eqn:RXA_Rdependence1}), we note the first term in Eq.~(\ref{eqn:RXA_Rdependence4}) cancels with the first term in Eq.~(\ref{eqn:RXA_Rdependence1}), and hence we recover Eq.~(\ref{eqn:RXA_Rdependence}).

\section{Gauge Covariance of Non-Abelian Berry Curvature}
In this section, we will prove the gauge covariance of the non-Abelian Berry curvature given in Eq.(83) in the main paper. We use the same notations as the main paper where we rotate our old set of basis states to a new set of basis states with a unitary matrix $\ket{\psi_{\overline{j}}} =\sum_{k}  \ket{\psi_{k}}\overline{ U}_{k\overline{j}}$.
Recall the definition of the non-Abelian Berry curvature (in the rotated basis), 
\begin{eqnarray}
\label{eqn:non_ab_BC}
    \overline{\bm\Omega}^{I\alpha J\beta} =    \frac{\partial  {\overline{\bm A}^{J\beta}}}{\partial R^{I\alpha}_{\rm n}} - 
     \frac{\partial  {\overline{\bm A}^{I\alpha}}}{\partial R^{J\beta}_{\rm n}} -
     \frac{i}{\hbar} \left[ {\overline{\bm A}^{I \alpha}}, {\overline{\bm A}^{ J \beta}}
    \right]
\end{eqnarray}
where
\begin{eqnarray}
     \overline{\bm A}^{J\beta} &=& \overline{\bm U}^{\dagger}{\bm A}^{J\beta}  \overline{\bm U} + i\hbar \overline{\bm U}^{\dagger}\nabla_{{J\beta}}\overline{\bm U}   \label{eqn:A_bar}
\end{eqnarray}
The derivative of the Berry connection in Eq.~(\ref{eqn:A_bar}) is then
\begin{eqnarray}
\label{eqn:dA_bar}
    \frac{\partial  {\overline{\bm A}^{J\beta}}}{\partial R^{I\alpha}_{\rm n}} = \overline{\bm U}^{\dagger} \frac{\partial  {{\bm A}^{J\beta}}}{\partial R^{I\alpha}_{\rm n}}\overline{\bm U} - \left[\overline{\bm U}^{\dagger}\nabla_{I\alpha}\overline{\bm U}  ,  \overline{\bm U}^{\dagger}{\bm A}^{J\beta}  \overline{\bm U}  \right] + i\hbar\nabla_{I\alpha} (\overline{\bm U}^{\dagger}\nabla_{{J\beta}}\overline{\bm U}   )
\end{eqnarray}
Use Eq.~(\ref{eqn:dA_bar}) the Abelian part of the non-Abelian Berry curvature in Eq.~(\ref{eqn:non_ab_BC}
) becomes
\begin{eqnarray}
\label{eqn:ab_BC1}
\frac{\partial  {\overline{\bm A}^{J\beta}}}{\partial R^{I\alpha}_{\rm n}} - 
     \frac{\partial  {\overline{\bm A}^{I\alpha}}}{\partial R^{J\beta}_{\rm n}} &=& \overline{\bm U}^{\dagger}\Big( \frac{\partial  {{\bm A}^{J\beta}}}{\partial R^{I\alpha}_{\rm n}} - \frac{\partial  {{\bm A}^{I\alpha}}}{\partial R^{J\beta}_{\rm n}} \Big)\overline{\bm U} -\left[\overline{\bm U}^{\dagger}\nabla_{I\alpha}\overline{\bm U}  ,  \overline{\bm U}^{\dagger}{\bm A}^{J\beta}  \overline{\bm U}  \right] \nonumber\\
\end{eqnarray}
Now evaluate the commutator in Eq.~(\ref{eqn:non_ab_BC}), we get
\begin{eqnarray}
\label{eqn:nab_BC1}
    -
     \frac{i}{\hbar} \left[ {\overline{\bm A}^{I \alpha}}, {\overline{\bm A}^{ J \beta}}\right] &=& -
     \frac{i}{\hbar} \overline{\bm U}^{\dagger}\left[ {{\bm A}^{I \alpha}}, {{\bm A}^{ J \beta}}\right]\overline{\bm U} + \left[ \overline{\bm U}^{\dagger}{\bm A}^{I\alpha}  \overline{\bm U}, \overline{\bm U}^{\dagger}\nabla_{J\beta}\overline{\bm U}\right] + \left[ \overline{\bm U}^{\dagger}\nabla_{I\alpha}\overline{\bm U}, \overline{\bm U}^{\dagger}{\bm A}^{J\beta}  \overline{\bm U}\right] \nonumber\\
     && + i\hbar  \left[ \overline{\bm U}^{\dagger}\nabla_{I\alpha}\overline{\bm U}, \overline{\bm U}^{\dagger}\nabla_{J\beta}\overline{\bm U}\right]
\end{eqnarray}
Adding Eq.~(\ref{eqn:ab_BC1}) and Eq.~(\ref{eqn:nab_BC1}) together, we get
\begin{align}
    \overline{\bm {\Omega}}^{I\alpha J\beta} =  \overline{\bm U}^{\dagger}\Big( \frac{\partial  {{\bm A}^{J\beta}}}{\partial R^{I\alpha}_{\rm n}} - \frac{\partial  {{\bm A}^{I\alpha}}}{\partial R^{J\beta}_{\rm n}} \Big)\overline{\bm U} -\frac{i}{\hbar} \overline{\bm U}^{\dagger}\left[ {{\bm A}^{I \alpha}}, {{\bm A}^{ J \beta}}\right]\overline{\bm U} = \overline{\bm U}^{\dagger}{\bm {\Omega}}^{I\alpha J\beta} \overline{\bm U}
\end{align} 
which validates the gauge covariance of the non-Abelian Berry cuvature in Eq.(83) in the main paper.

\section{Momentum Conservation with Fully Adiabatic Representation}
\subsection{Hamiltionian and Equations of Motion}
Given a complete nuclear-electronic Hamiltonian,
\begin{eqnarray}
    \label{eqn:H_diab}
      \tilde{\bm H} &=& \sum_{I}\frac{\hat{{ \bm P}}_{\rm n}^{I^2}}{2M_{I}}+  \tilde{\bm V},
\end{eqnarray}
it is quite standard to 
construct a diabatic-to-adiabatic transformation and rewrite the Hamiltonian in this new basis:
\begin{equation}
\begin{split}
        \hat{\bm H} &= {\bm U}^{\dagger}\tilde{\bm H}{\bm U}\\
        & = \sum_{I}\frac{-\hbar^2}{2M_{I}}{\bm U}^{\dagger} \nabla^{I^2}_{\rm n}  {\bm U}  + {\bm U}^{\dagger}\tilde{\bm V}  {\bm U}  \\
        & = \sum_{I}\frac{-\hbar^2}{2M_{I}}( {\bm U}^{\dagger} \nabla^{I}_{\rm n} {\bm U})({\bm U}^{\dagger} \nabla^{I}_{\rm n} {\bm U} )-   \sum_{I}\frac{i\hbar}{M_{I}}{\bm U}^{\dagger} \nabla^{I}_{\rm n}  {\bm U}  \nabla^{I}_{\rm n} - \frac{\hbar^2\nabla_{\rm n}^{I^2}}{2M_{I}}+ \hat{\bm V} \\
        & = \sum_{I}\frac{(-i\hbar\nabla_{\rm n}^{I}-\bm A^{I})^2}{2M_{I}} +\hat{\bm V}
\end{split}
\end{equation}
If we now make the quantum-classical approximation, $\hat{{ \bm P}}_{\rm n} \rightarrow \bm P_{\rm n}$, we obtain what we call ``a fully adiabatic Hamiltonian" and the corresponding energy expression from the main text (Eqs. (84) and (85)):
\begin{eqnarray}
    \hat{H} &=& \sum_{I}\frac{(\bm P_{\rm n}^{I}-\bm A^{I})^2}{2M_{I}} +\hat{\bm V} \\
    E &=& \sum_{I}\frac{{ \bm P}_{\rm n}^{I^2}}{2M_{I}}
    +\mbox{Tr}\left[\hat{\bm \sigma} \left(\sum_{I}\frac{{-2\bm P_{\rm n}^{I}} \cdot  {{\bm A}^{I}} + \bm A^{I^2} }{2M_{I}} + \hat{\bm V}\right)\right]
\end{eqnarray}

In the spirit of a semiclassical calculation, we may then derive the corresponding equations of motion for the electronic density matrix (from the electronic Schrodinger equation) and the nuclear degrees of freedom (from Hamilton's equations):
\begin{eqnarray}
    \label{eqn:d-dot-adiab}
\dot{\hat{\bm\sigma}} &=&  -\frac{i}{\hbar} \left[ \hat{\bm V} + \sum_{I}
\frac{{ -2\bm P_{\rm n}^{I}} \cdot  {{\bm A}^{I}} + \bm A^{I^2} }{2M_{I}}, \hat{\bm\sigma} \right] \\
\dot{R}^{I\alpha}_{\rm n} &=& \frac{\partial E}{\partial P^{I\alpha}_{\rm n}} =\frac{P^{I\alpha}_{\rm n}-\mbox{Tr}\left( \hat{\bm\sigma}\bm A^{I\alpha}\right)}{M_I}  \label{eqn:R-dot-adiab} \\
        \dot{P}^{I\alpha}_{\rm n} &=&  -\frac{\partial E}{\partial R^{I\alpha}_{\rm n}}=-\sum_{J\beta} \mbox{Tr}\left(\hat{\bm\sigma} \left( \frac{\partial \hat{\bm V}}{\partial R^{I\alpha}_{\rm n}}
 -\frac{P^{J\beta}_{\rm n}}{M_J} \frac{\partial {\bm A^{J\beta}}}{\partial R^{I\alpha}_{\rm n}}
 + \frac{\bm A^{J\beta}} {2M_J} \frac{\partial {\bm A^{J\beta}}}{\partial R^{I\alpha}_{\rm n}} +\frac{\partial {\bm A^{J\beta}}}{\partial R^{I\alpha}_{\rm n}} \frac{\bm A^{J\beta}} {2M_J} 
 \right) \right)\label{eqn:P-dot-adiab} 
\end{eqnarray}
Similarly, we can introduce the kinetic nuclear momentum $\bm \pi^{I\alpha}_{\rm n} = P^{I\alpha}_{\rm n}-\mbox{Tr}\left( \hat{\bm \sigma}{\bm A}^{I\alpha}\right) $.
\subsection{Linear Momentum Conservation}
Let us now evaluate the total linear momentum:
\begin{eqnarray}
\label{eqn:Ptot_dot2}
\dot {\bm P}_{\rm tot}^{\alpha} &=&  \sum_{I} \dot{\pi}_{\rm n}^{I\alpha} + \mbox{Tr}\left( \dot{\hat{\bm\sigma}}  {{\bm P}_{\rm e}^{\alpha}} \right)  + \mbox{Tr}\left(  {\hat{\bm\sigma}}  
    \dot {{\bm P}}_{\rm e}^{\alpha} \right) \\
    &=& \sum_{I} \dot{P}_{\rm n}^{I\alpha} + \mbox{Tr}\Big[ \dot{\hat{\bm\sigma}} ( {{\bm P}_{\rm e}^{\alpha} - \sum_{I}{\bm A}^{I\alpha}} )\Big]  + \mbox{Tr}\Big[   {\hat{\bm\sigma}}  
    ( \dot{\bm P}_{\rm e}^{\alpha} - \sum_{I}{\dot{\bm A}^{I\alpha}} )\Big]  \label{eqn:Ptot_simply1} \\
    & =& \sum_I \dot{P}_{\rm n}^{I\alpha} + \sum_{jk}\dot{\hat{\sigma}}_{kj} \xi^{\alpha}_{k}\delta_{jk}+\sum_{IJjk\beta}\dot{R}^{J\beta}_{\rm n}\hat{\sigma}_{kj} {\xi}^{J\beta, I\alpha}_{k}\delta_{jk}\label{eqn:Ptot_simply2}
\end{eqnarray}
To go from Eq.~(\ref{eqn:Ptot_simply1}) to Eq.~(\ref{eqn:Ptot_simply2}) here, we used  Eqs. (75) and (76) in the main manuscript. Now we evaluate each term in Eq.~(\ref{eqn:Ptot_simply2}).
\begin{itemize}
    \item To simplify the first term $\sum_I \dot{P}_{\rm n}^{I\alpha}$ in Eq.~(\ref{eqn:Ptot_simply2}), we use Eq.~(\ref{eqn:vanish}) and Eq.~(\ref{eqn:P-dot-adiab}),
\begin{eqnarray}
        \sum_I \dot{P}_{\rm n}^{I\alpha} &=& \sum_{IJ\beta jkl} \hat{\sigma} _{kj}\Bigg[
 \frac{P^{J\beta}_{\rm n}}{M_J}\Big(- \xi^{J\beta,I\alpha}_{k}\delta_{jk}  +\frac{i}{\hbar}(\xi^{I\alpha}_{k}-\xi^{I\alpha}_{j})A^{J\beta}_{jk} \Big)\nonumber
 \\
 &&+\frac{ A^{J\beta}_{jl}} {2M_J} \Big( \xi^{J\beta,I\alpha}_{k}\delta_{lk}  -\frac{i}{\hbar} (\xi^{I\alpha}_{k}-\xi^{I\alpha}_{l})A^{J\beta}_{lk} \Big) \label{eqn:Pn_dot_1}\nonumber\\
&&+\Big( \xi^{J\beta,I\alpha}_{l}\delta_{jl}  -\frac{i}{\hbar} (\xi^{I\alpha}_{l}-\xi^{I\alpha}_{j})A^{J\beta}_{jl} \Big) \frac{ A^{J\beta}_{lk}} {2M_J} 
 \Bigg] \\
 &=& \sum_{IJ\beta jkl} \hat{\sigma} _{kj}\Bigg[-
 \frac{P^{J\beta}_{\rm n}}{M_J} \xi^{J\beta,I\alpha}_{k}\delta_{jk}  + \frac{  A^{J\beta}_{jk}} {2M_J} (\xi^{J\beta,I\alpha}_{k} + \xi^{J\beta,I\alpha}_{j}) \nonumber\\
 && + \frac{i}{2M_{J}\hbar} (\xi^{I\alpha}_{k}-\xi^{I\alpha}_{j}) ( 2P^{J\beta}_{\rm n} A^{J\beta}_{jk} - A^{J\beta}_{jl} A^{J\beta}_{lk} )\label{eqn:Pn_dot_2} \Bigg] 
\end{eqnarray}
Here we recognize $-\sum_{I}\mbox{Tr}\left(\hat{\sigma} \frac{\partial \hat{\bm V}}{\partial R^{I\alpha}_{\rm n}}\right)=0 $ for a translation-invariant potential energy surface. 
\item We can evaluate the second term in Eq.~(\ref{eqn:Ptot_simply2}) by plugging in Eq.~(\ref{eqn:d-dot-adiab}) ,
\begin{eqnarray}
     \sum_{jk}\dot{\hat{\sigma}}_{kj} \xi^{\alpha}_{k}\delta_{jk} =  -\frac{i}{\hbar} \sum_{IJ\beta jkl} \frac{1}{2M_{J}} \hat{\sigma}_{kj}(\xi^{I\alpha}_{k}-\xi^{I\alpha}_{j}) ( 2\dot{P}^{J\beta}_{\rm n} A^{J\beta}_{jk} - A^{J\beta}_{jl} A^{J\beta}_{lk} )\label{eqn:Ptot_simply3}
\end{eqnarray}
\item  We can evaluate the third term in Eq.~(\ref{eqn:Ptot_simply2}) by plugging in Eq.~(\ref{eqn:R-dot-adiab}):
\begin{eqnarray}
\label{eqn:Ptot_simply4}
    \sum_{IJjk\beta}\dot{R}^{J\beta}_{\rm n}\hat{\sigma}_{kj} {\xi}^{J\beta, I\alpha}_{k}\delta_{jk} =   \sum_{IJjk\beta}  \frac{  \hat{\sigma} _{kj} } {M_J}\Big[P^{J\beta}_{\rm n} -{\rm Tr}(\bm \sigma \bm A^{J\beta})\Big]\xi^{J\beta,I\alpha}_{k} \delta_{jk}
\end{eqnarray}
\end{itemize}

Use Eqs.~(\ref{eqn:Pn_dot_2})-(\ref{eqn:Ptot_simply4}), we can evaluate the time-dependence of the total linear momentum in Eq.~(\ref{eqn:Ptot_simply2}). We first note that the first term in Eq.~(\ref{eqn:Pn_dot_2}) cancels with the first term in Eq.~(\ref{eqn:Ptot_simply4}). We further note the last term in Eq.~(\ref{eqn:Pn_dot_2}) cancels with Eq.~(\ref{eqn:Ptot_simply3}). The remaining terms in Eq.~(\ref{eqn:Ptot_simply2}) are, 
\begin{eqnarray}
    \dot {\bm P}_{\rm tot}^{\alpha} &=& - \sum_{IJ\beta jk} \frac{ \hat{\sigma} _{kj} } {M_J}\Bigg[ {\rm Tr}[\bm \sigma \bm A^{J\beta}]\xi^{J\beta,I\alpha}_{k} \delta_{jk}- A^{J\beta}_{jk}\frac{1}{2}(\xi^{J\beta,I\alpha}_{k} + \xi^{J\beta,I\alpha}_{j})\Bigg] \label{eqn:Ptot_dot}
\end{eqnarray}
From Eq.~(\ref{eqn:Ptot_dot}), we see that the total linear momentum with the fully adiabatic equations of motions is generally not conserved -- unless one is able to implement the  phase convention  $\xi_{k} = \xi_{j}$ = 0 (which is difficult for on the fly calculations as discussed in the main text).

\subsection{Angular Momentum Conservation}
Next, we perform a similar derivation for total angular momentum conservation, beginning with Eqs. (78)-(79) in the main text:
\begin{eqnarray}
\dot {\bm J}_{\rm tot}^{\alpha} &=&  \sum_{I\eta\gamma}\epsilon_{\alpha\eta\gamma}  R^{I\eta}_{\rm n} \dot  \pi^{I\gamma}_{\rm n}  +   \mbox{Tr}\left( \hat{\bm\sigma}  \dot {\bm J}_{\rm e}^{\alpha}  + \dot {\hat{\bm \sigma}}   {\bm J}_{\rm e}^{\alpha} \right)\\
&=&  \sum_{I\eta\gamma}\epsilon_{\alpha\eta\gamma}  R^{I\eta}_{\rm n} \dot  P^{I\gamma}_{\rm n} + \mbox{Tr}\left[\hat{\bm\sigma}  \left( \dot {\bm J}_{\rm e}^{\alpha} -\sum_{I\eta\gamma}\epsilon_{\alpha\eta\gamma}  R^{I\eta}_{\rm n}\dot{{\bm A}}^{I\gamma} \right)\right] \nonumber\\
&&+
\mbox{Tr}\left[\dot{\hat{\bm\sigma}}   \left({\bm J}_{\rm e}^{\alpha} -\sum_{I\eta\gamma}\epsilon_{\alpha\eta\gamma}  R^{I\eta}_{\rm n}{{\bm A}}^{I\gamma} \right)\right]
\label{eqn:Jtot_dot2}
\end{eqnarray}
We now evaluate each term in Eq.~(\ref{eqn:Jtot_dot2}).
\begin{itemize}
    \item From Eq.~(\ref{eqn:P-dot-adiab}), we can write the first term in Eq.~(\ref{eqn:Jtot_dot2}),
\begin{eqnarray}
     \sum_{I\eta\gamma}\epsilon_{\alpha\eta\gamma}  R^{I\eta}_{\rm n} \dot  P^{I\gamma}_{\rm n} &=& \sum_{IJ\beta\eta\gamma jkl} \epsilon_{\alpha\eta\gamma}  R^{I\eta}_{\rm n} \hat{\sigma} _{kj}\Bigg(
 \frac{P^{J\beta}_{\rm n}}{M_J} \frac{\partial { A^{J\beta}_{jk}}}{\partial R^{I\gamma}_{\rm n}}
 - \frac{ A^{J\beta}_{jl}} {2M_J} \frac{\partial A^{J\beta}_{lk}}{\partial R^{I\gamma}_{\rm n}} - \frac{\partial A^{J\beta}_{jl}}{\partial R^{I\gamma}_{\rm n}} \frac{ A^{J\beta}_{lk}} {2M_J} \Bigg)\label{eqn:Jn_dot}
\end{eqnarray}
Here we recognize that $\sum_{I\eta\gamma}\epsilon_{\alpha\eta\gamma}  R^{I\eta}_{\rm n} \mbox{Tr}\left(\hat{\sigma} \frac{\partial \hat{\bm V}}{\partial R^{I\gamma}_{\rm n}}\right)=0 $ due to isotropy of space. For the last two terms on the right-hand side of Eq.~(\ref{eqn:Jn_dot}), we can use Eq.~(\ref{eqn:RXA_Rdependence}) so that the last terms in Eq.~(\ref{eqn:Jn_dot}) become
\begin{eqnarray}
    && -\sum_{IJ\beta\eta\gamma jkl} \epsilon_{\alpha\eta\gamma}  R^{I\eta}_{\rm n} \hat{\sigma} _{kj}\Bigg(
 \frac{ A^{J\beta}_{jl}} {2M_J} \frac{\partial A^{J\beta}_{lk}}{\partial R^{I\gamma}_{\rm n}} + \frac{\partial A^{J\beta}_{jl}}{\partial R^{I\gamma}_{\rm n}} \frac{ A^{J\beta}_{lk}} {2M_J} \Bigg) \nonumber\\
     &=&-\sum_{IJ\beta\eta\gamma jkl} \hat{\sigma} _{kj}\Bigg\{\frac{ A^{J\beta}_{jl}} {2M_J} \Bigg(  \epsilon_{\alpha\eta\gamma} R^{I\eta}_{\rm n} \Big[\frac{i}{\hbar}(\xi^{I\gamma}_{k} - \xi^{I\gamma}_{l})A^{J\beta}_{lk} - \delta_{lk}\xi^{J\beta,I\gamma}_{k}\Big]
    - \epsilon_{\alpha\beta\gamma} (A_{lk}^{J\gamma} + \xi^{J\gamma}_{k}\delta_{lk} )
     \Bigg) \nonumber
     \\&&+  \Bigg( \epsilon_{\alpha\eta\gamma} R^{I\eta}_{\rm n} \Big[\frac{i}{\hbar}(\xi^{I\gamma}_{l} - \xi^{I\gamma}_{j})A^{J\beta}_{jl} - \delta_{jl}\xi^{J\beta,I\gamma}_{l}\Big] - \epsilon_{\alpha\beta\gamma} (A_{jl}^{J\gamma} +  \xi^{J\gamma}_{l}\delta_{jl} )
     \Bigg)\frac{ A^{J\beta}_{lk}} {2M_J} \Bigg\} \\
     & = & -\sum_{IJ\beta\eta\gamma jkl} \hat{\sigma} _{kj}\epsilon_{\alpha\eta\gamma} R^{I\eta}_{\rm n} \Big[\frac{i}{\hbar}(\xi^{I\gamma}_{k} - \xi^{I\gamma}_{j})\frac{ A^{J\beta}_{jl}} {2M_J}A^{J\beta}_{lk} -\frac{ A^{J\beta}_{jk}} {2M_J}(\xi^{J\beta,I\gamma}_{k} + \xi^{J\beta,I\gamma}_{j} )\Big]\nonumber \\
     &&+ \sum_{J\beta\gamma jkl} \frac{1}{2M_J}\hat{\sigma} _{kj}\epsilon_{\alpha\beta\gamma}\Big[A^{J\gamma}_{jl} A^{J\beta}_{lk} + A^{J\beta}_{jl} A^{J\gamma}_{lk}  + A^{J\beta}_{jk} (\xi^{J\gamma}_{k} + \xi^{J\gamma}_{j} )\Big]
\label{eqn:Jn_dot2}
\end{eqnarray}

\item If we rearrange Eq. (45) in the main paper, the second term in Eq.~(\ref{eqn:Jtot_dot2}) becomes
\begin{eqnarray}
    \label{eqn:Je_dot_minus_A_dot}
    &&\mbox{Tr}\left(\hat{\bm\sigma}  \dot {\bm J}_{\rm e}^{\alpha} \right)-\sum_{I\eta\gamma}\epsilon_{\alpha\eta\gamma}  R^{I\eta}_{\rm n}\mbox{Tr}\left( \hat{\bm\sigma} \dot{{\bm A}}^{I\gamma} \right) \nonumber\\
    &=& \sum_{IJ\eta\beta\gamma jk}\epsilon_{\alpha\eta\gamma}R^{I\eta}_{\rm n} \dot{R}^{J\beta}_{\rm n}\hat{\sigma}_{kj}\left[-\frac{\partial {A}^{J\beta}_{jk}}{\partial R^{I\gamma}_{\rm n}}+ \frac{i}{\hbar}(\xi^{I\gamma}_{k} - \xi^{I\gamma}_{j})A^{J\beta}_{jk}\right] \\
    &=&\sum_{IJ\eta\beta\gamma jk}\frac{1}{M_{J}}\epsilon_{\alpha\eta\gamma}R^{I\eta}_{\rm n} \hat{\sigma}_{kj}\Big[{P}^{J\beta}_{\rm n}  -{\rm Tr}(\bm \sigma \bm A^{J\beta})\Big]\left[-\frac{\partial {A}^{J\beta}_{jk}}{\partial R^{I\gamma}_{\rm n}}+ \frac{i}{\hbar}(\xi^{I\gamma}_{k} - \xi^{I\gamma}_{j})A^{J\beta}_{jk}\right] \label{eqn:Je_dot_minus_A_dot3}
\end{eqnarray}
Now we plug in Eq.~(\ref{eqn:RXA_Rdependence}) for those terms in Eq.~(\ref{eqn:Je_dot_minus_A_dot3}) that depend on ${\rm Tr}(\bm \sigma \bm A^{J\beta})$, and find:
\begin{eqnarray}
     &&\sum_{IJ\eta\beta\gamma jk}\epsilon_{\alpha\eta\gamma}R^{I\eta}_{\rm n} \frac{{\mbox Tr}(\hat{\bm\sigma}\bm A^{J\beta})}{M_{J}}\hat{\sigma} _{kj}\left[\frac{\partial {A}^{J\beta}_{jk}}{\partial R^{I\gamma}_{\rm n}}- \frac{i}{\hbar}(\xi^{I\gamma}_{k} - \xi^{I\gamma}_{j})A^{J\beta}_{jk}\right]\nonumber \\
     &=& \sum_{IJ\eta\beta\gamma jk} \frac{{\mbox Tr}(\hat{\bm\sigma}\bm A^{J\beta})}{M_{J}} \hat{\sigma} _{kj}\Bigg[-\epsilon_{\alpha\eta\gamma}R^{I\eta}_{\rm n}\delta_{jk}\xi^{J\beta,I\gamma}_{k} -\epsilon_{\alpha\beta\gamma}(A^{J\gamma}_{jk}+\xi^{J\gamma}_{k}\delta_{jk})\Bigg]
\end{eqnarray}

\item Lastly, we can evaluate the third term in Eq. (\ref{eqn:Jtot_dot2}):
\begin{eqnarray}   
&&\mbox{Tr}\left(\dot{\hat{\bm\sigma}}  {\bm J}_{\rm e}^{\alpha} \right)-\sum_{I\eta\gamma}\epsilon_{\alpha\eta\gamma}  R^{I\eta}_{\rm n}\mbox{Tr}\left(  \dot{\hat{\bm\sigma}}{{\bm A}}^{I\gamma} \right)\nonumber \\&=& \sum_{I\eta\gamma jk} \epsilon_{\alpha\eta\gamma}R^{I\eta}_{\rm n}\dot{\hat{\bm \sigma}}_{kj}\xi^{I\gamma}_{k}\delta_{jk} 
\label{eq:joe:int}
\\
& = & - \sum_{IJ\beta\eta\gamma jkl} \epsilon_{\alpha\eta\gamma}R^{I\eta}_{\rm n} \frac{i}{2M_{J}\hbar} \hat{\sigma}_{kj}(\xi^{I\gamma}_{k}-\xi^{I\gamma}_{j}) ( 2P^{J\beta}_{\rm n} A^{J\beta}_{jk} - A^{J\beta}_{jl} A^{J\beta}_{lk} )\label{eqn:Je_minus_A} 
\end{eqnarray}
\noindent Here, we have used  Eq. (46) in the main paper to recover Eq. (\ref{eq:joe:int}) and Eq.~(\ref{eqn:Ptot_simply3}) to recover Eq. (\ref{eqn:Je_minus_A}).
\end{itemize}

With Eqs.~(\ref{eqn:Jn_dot})-(\ref{eqn:Je_minus_A}), we can now evaluate the total time-dependence of the angular momentum in Eq.~(\ref{eqn:Jtot_dot2}). We first note that together with 
 the first term in Eq.~(\ref{eqn:Je_minus_A}) and the first term in Eq.~(\ref{eqn:Jn_dot}) cancel the terms that depend on $P_{\rm n} $ in Eq.~(\ref{eqn:Je_dot_minus_A_dot3}). We further see the second term in Eq.~(\ref{eqn:Je_minus_A}) cancel with the first term in Eq.~(\ref{eqn:Jn_dot2}). The remaining terms in Eq.~(\ref{eqn:Jtot_dot2}) are then
\begin{eqnarray}
    \dot {\bm J}_{\rm tot}^{\alpha} &=& -\sum_{IJ\beta \eta\gamma jk}\frac{ \hat{\sigma} _{kj} } {M_J}\epsilon_{\alpha\eta\gamma} R^{I\eta}_{\rm n}\Bigg[ {\rm Tr}[\bm \sigma \bm A^{J\beta}]\xi^{J\beta,I\gamma}_{k} \delta_{jk}- A^{J\beta}_{jk}\frac{1}{2}(\xi^{J\beta,I\gamma}_{k} + \xi^{J\beta,I\gamma}_{j})\Bigg] \nonumber\\
    && + \sum_{J\beta\gamma jkl} \frac{ 1} {2M_J} \hat{\sigma} _{kj}\epsilon_{\alpha\beta\gamma}\Bigg[A^{J\gamma}_{jl} A^{J\beta}_{lk} + A^{J\beta}_{jl} A^{J\gamma}_{lk} -2\mbox{Tr}( \hat{\bm\sigma}{{\bm A}}^{J\beta})A^{J\gamma}_{jk}  \nonumber\\
    &&- 2\mbox{Tr}(\hat{\bm\sigma}{{\bm A}}^{J\beta} )\xi^{J\gamma}_{k}\delta_{jk} + A^{J\beta}_{jk}(\xi^{J\gamma}_{k} + \xi^{J\gamma}_{j} )\Bigg] \label{eqn:Jtotal1}\\
    & = & -\sum_{IJ\beta \eta\gamma jk}\frac{ \hat{\sigma} _{kj} } {M_J}\epsilon_{\alpha\eta\gamma} \Bigg\{ R^{I\eta}_{\rm n}\Bigg[ {\rm Tr}(\bm \sigma \bm A^{J\beta})\xi^{J\beta,I\gamma}_{k} \delta_{jk}- A^{J\beta}_{jk}\frac{1}{2}(\xi^{J\beta,I\gamma}_{k} + \xi^{J\beta,I\gamma}_{j})\Bigg] \nonumber\\
    &&+  \mbox{Tr}( \hat{\bm\sigma}{{\bm A}}^{J\eta} )\xi^{J\gamma}_{k}\delta_{jk} -   A^{J\eta}_{jk}\frac{1}{2}(\xi^{J\gamma}_{k} + \xi^{J\gamma}_{j} )\Bigg\}\label{eqn:Jtotal2}
\end{eqnarray}
To go from Eq.~(\ref{eqn:Jtotal1}) to Eq.~(\ref{eqn:Jtotal2}), we recognize $\sum_{\beta\gamma}\epsilon_{\alpha\beta\gamma}\Big(\bm A^{J\gamma} \bm A^{J\beta}+ \bm A^{J\beta}\bm A^{J\gamma} \Big) = 0 $ and \\
$\sum_{\beta\gamma}\epsilon_{\alpha\beta\gamma} \mbox{Tr}( \hat{\bm\sigma}{{\bm A}}^{J\beta})\mbox{Tr}( \hat{\bm\sigma}{{\bm A}}^{J\gamma}) = 0$. 

Eq.~(\ref{eqn:Jtotal2}) cannot be simplified further. As for the case of linear momentum, we conclude that the total angular momentum with the fully adiabatic equations of motions is generally not conserved--unless one is able to implement the phase convention $\xi_{k} = \xi_{j}$ = 0. 


\section{Methoxy Radical Ehrenfest Dynamics}

\subsection{Cartesian Coordinates}
The initial geometry used to propagate Ehrenfest dynamics for the vibrational motion of the methoxy radical was optimized with the hydrogens on the carbon at the unrestricted Hartree-Fock level of theory  with 6-31G basis set. The Cartesian coordinates are given in Angstrom here. 
\begin{table}[H]  
\begin{tabular}{c c c c} 
\hline\hline
    C        & 0.0036502509   & 0.0000000000   &-0.0152491302 \\
    O        &-0.0055615054   &0.0000000000   & 1.4089433974 \\
    H        & 1.0467953363  & 0.0000000000   &-0.3112007981 \\
    H       & -0.4714567426  &-0.8895741740   &-0.4027895416 \\
    H       & -0.4714567426  & 0.8895741740   &-0.4027895416 \\
    \hline
\end{tabular} 
\end{table}

For the rotation of the methoxy radical, the intial geometry was further rotated so that the C-O bond is align at the x-axis and the carbon atom is at the origin.

\subsection{Trajectory Details}
\subsubsection{Vibration of the Methoxy Radical}
\begin{figure}[h]
\includegraphics[width=0.8\textwidth]{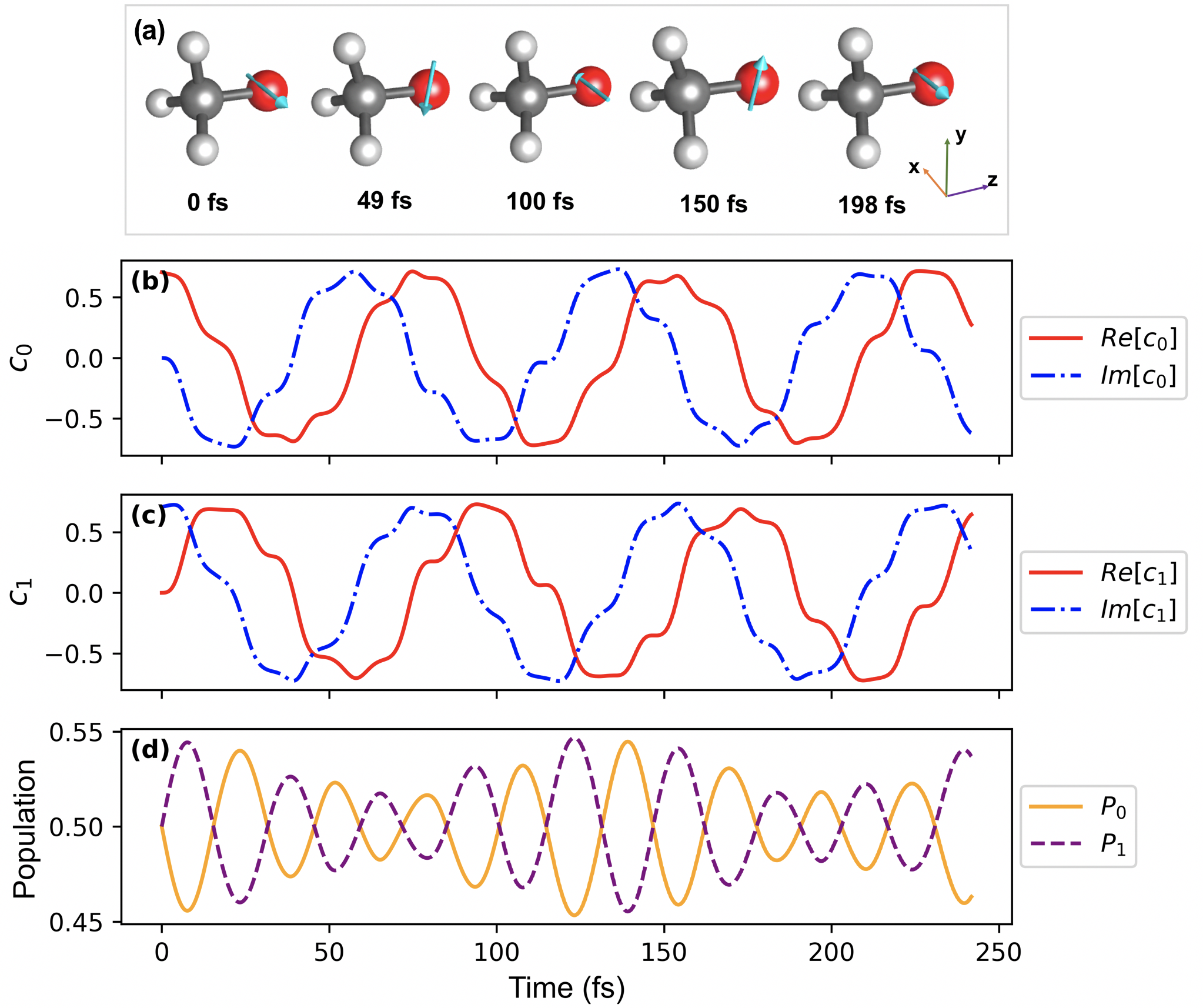} 
\caption{ The plot of (a) the Ehrenfest averaged atom-based magnetic dipole moments, (b) real (red solid line) and imaginary (blue dotted-dash line) values for the amplitude of state 0, (c) real (red solid line) and imaginary (blue dotted-dash line) values for the amplitude of state 1, and (d) the population for state 0 (brown solid line) and state 1 (purple dash line) with time. For motion along this vibration, Ehrenfest dynamics predict that the spin on the O atom should vary in time as the quantum wavefunction oscillates between populating the $\ket{GHF}$ and $\ket{TGHF}$ states.} \label{fig:pop}
\end{figure}

We performed Ehrenfest dynamics in a basis generated with a generalized Hartree-Fock (GHF) caclulation accounting for spin-orbit coupling (SOC); by hand, we generated a time-reversed GHF (TGHF) state. The basis set used was 6-31G basis set. The direction of the initial velocity vector corresponds to the lowest normal mode (shown in Fig.1(a) in the main manuscript), with an initial kinetic energy of 0.005 a.u. ($\approx 1491K$). A step size of 5 a.u. (0.121 fs) was used.  The initial amplitudes for the Kramers doublet pair were $\bm  c = (\frac{1}{\sqrt{2}}, \frac{i}{\sqrt{2}})$ in the $\left\{\ket{GHF}, \ket{TGHF}\right\}$ basis. 
In Fig. 1(d) in the main manuscript, we showed the initial spin magnetic moments for each atom (at time zero) appropriate for this initial wavefunction. Now, in Fig.~\ref{fig:pop}(a), we plot how the spin magnetic moments rotate in the xy plane as a function of time. Interestingly, for this vibrational motion (with little or no rotation), the spin magnetic moments for $\ket{GHF}$ and $\ket{TGHF}$ states do not change much with time from their initial values (which were plotted in Fig.1(b)-(c) in the main manuscript). 
And yet, according to Fig.~\ref{fig:pop}(b)-(d),  the amplitudes and population change frequently with time during the Ehrenfest trajectory, consistent with the fact that the spin moment on the O atom is oscillating with a period of about 200fs. 

\begin{figure}[h] 
\includegraphics[width=0.8\textwidth]{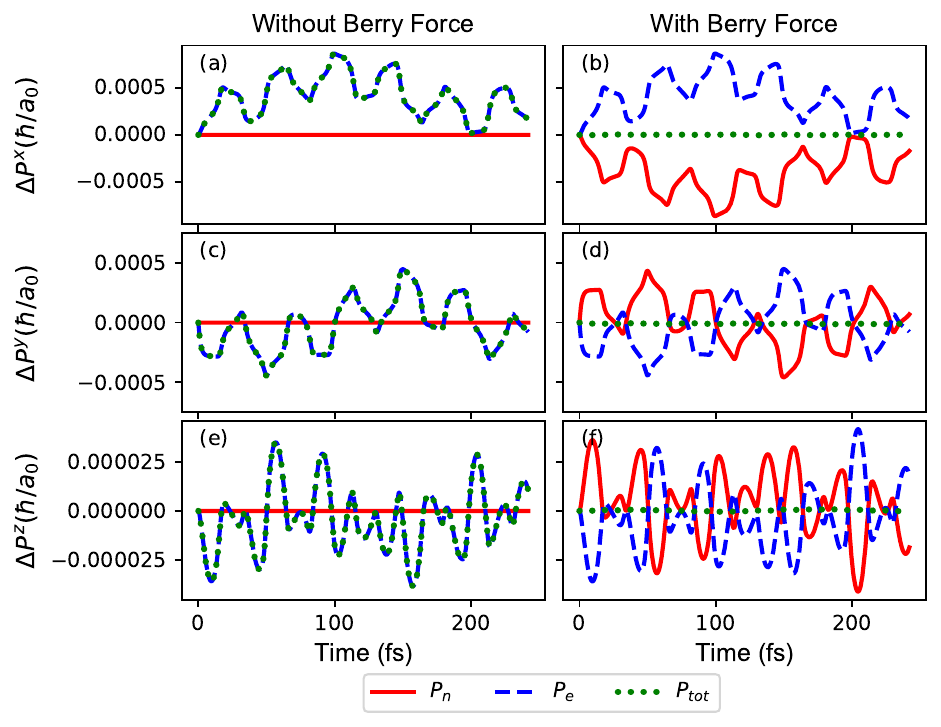}
\caption{ The change in the real-time  linear momentum relative to time zero for motion along the lowest-normal-mode vibration of the methoxy radical according to Ehrenfest dynamics without (Eqs. (17)-(19)) or  with (Eqs. (68)-(71)) a Berry force. The nonzero initial values of the linear momentum are:  $P^{x}_{\rm e} = -0.0004 \hbar/a_{\rm 0}$ and $P^{y}_{\rm n} = 0.0221\hbar/a_{\rm 0}$.  Results are shown for all three Cartesian coordinates, $x$, $y$ and $z$.  In Figs. (b), (d), (f), with Berry force, the nuclear (red solid line) and electronic (blue dashed line) linear momentum transfer between each other, and the total linear momentum (green dotted line) is conserved. By contrast, without Berry force [as shown in Fig.(a), (c) and (e)], the total linear momentum (green dotted line) is not conserved.}\label{fig:mom}
\end{figure}

In Fig.~\ref{fig:mom} (a)(c)(e) and (b)(d)(f), we plot the change in nuclear (red solid line), electronic (blue dashed line), and total linear momentum (green dotted line) without Berry force and with Berry force, respectively. We first consider the case when Berry force is not included and standard Ehrenfest equations of motion are used. (Eqs. (17)-(19) in the main manuscript). In Fig.~\ref{fig:mom} (a)(c)(e), when Berry force is not included, the nuclear linear momentum is calculated from the canonical momentum $\bm P_{\rm n}$, which is conserved for a translationally invariant surface, and hence there is no change in the nuclear momentum (red solid line). However, the electronic linear momentum (blue dashed line) changes with time and hence the total linear momentum (green dotted line) is not conserved. 

Next, let us consider the case when  Berry force is included and the effective Ehrenfest equations of motion (Eqs. (68)-(71) in the main manuscript) are propagated. In Fig.~\ref{fig:mom} (b)(d)(f), we see the nuclear linear momentum $\bm \pi_{\rm n}$ (Eq. (67) in the main manuscript) now changes with time. The electronic linear momentum  (blue dashed line) also changes with time, but in the opposite direction so that  the total linear momentum is conserved.

\subsubsection{Rotation of the Methoxy Radical}
\begin{figure}[h] 
\includegraphics[width=0.8\textwidth]{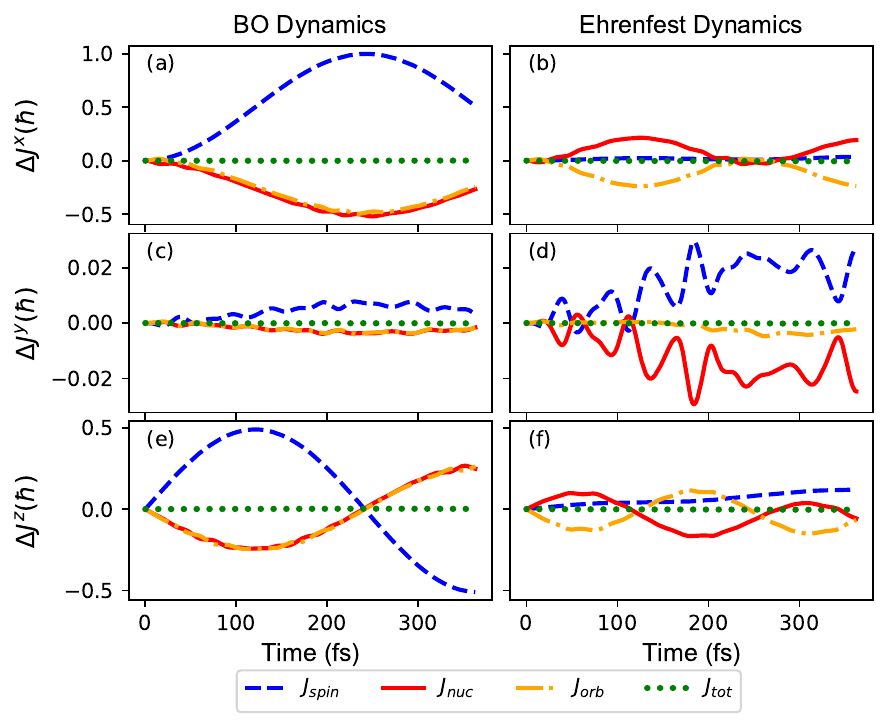}
\caption{ The change in the real-time angular momentum relative to time zero according to  BO+Berry force dynamics ((a), (c), (e)) versus Ehrenfest+Berry force dynamics ((b),(d) and (f)) during a rotation of the methoxy radical.    The nonzero initial value components of the angular momentum are: $J^{x}_{\rm orb} = 0.2373 \hbar$, $L^{z}_{\rm orb} = -0.0044 \hbar$,  $J^{y}_{\rm nuc} = 39.28 \hbar$, $J^{x}_{\rm spin} = -0.4999 \hbar$, and $J^{z}_{\rm spin} = 0.0097 \hbar$.  We plot the angular momentum along all  three Cartesian coordinates $x$, $y$ and $z$. Because the Berry force is included in both approaches, the total linear momentum (green dotted line) is always conserved in all the subplots. For our choice of BO state (that rotates with the molecule), as shown in  panels (a), (c), (e), the overall spin changes far too fast. By contrast, as shown in panels (b),(d) and (f), the Ehrenfest approach correctly slows down the spin dynamics.} \label{fig:angmom_bo_ehren}
\end{figure}

In Fig. 3(a) in the main manuscript, we performed Born-Oppenheimer (BO) dynamics and Ehrenfest dynamics with the GHF+SOC approach and 6-31G basis set when the direction of the initial velocity vector corresponded to an impulsive rotational force around the y axis. For those calculations, the initial nuclear angular momentum was $J_{\rm nuc} = (0.0, 39.28, 0.0)$ with an initial kinetic energy of 0.006 a.u. ($\approx 1948K$). A step size of 10 a.u. (0.242 fs) was used.  The initial amplitudes for the Kramers doublet pair were set $\bm  c = (1,0)$. In Fig. \ref{fig:angmom_bo_ehren}(a)(c)(e) and (b)(d)(f), we plot the nuclear (red solid line), electronic orbital (orange dashed-dotted line), spin (blue dashed line), and total (green dotted line) angular momentums changes with time as obtained with the BO and effective Ehrenfest approach (Eqs. (68)-(71) in the main manuscript), respectively. Because both approaches include Berry force, the total angular momentum (green dotted line) is conserved for both the BO and Ehrenfest dynamics. Comparing Fig \ref{fig:angmom_bo_ehren}(a)(e) and (b)(f), we see the change in the electronic spin angular momentum (blue dashed line) in the Ehrenfest approach is much smaller than in the BO approach. As discussed in the main manuscript, within the BO approximation, the electronic spin responds too fast (instantaeously) to the molecular rotation. Ehrenfest, on the other hand, correctly slows down the spin change. 

\begin{figure}[h] 
\includegraphics[width=0.8\textwidth]{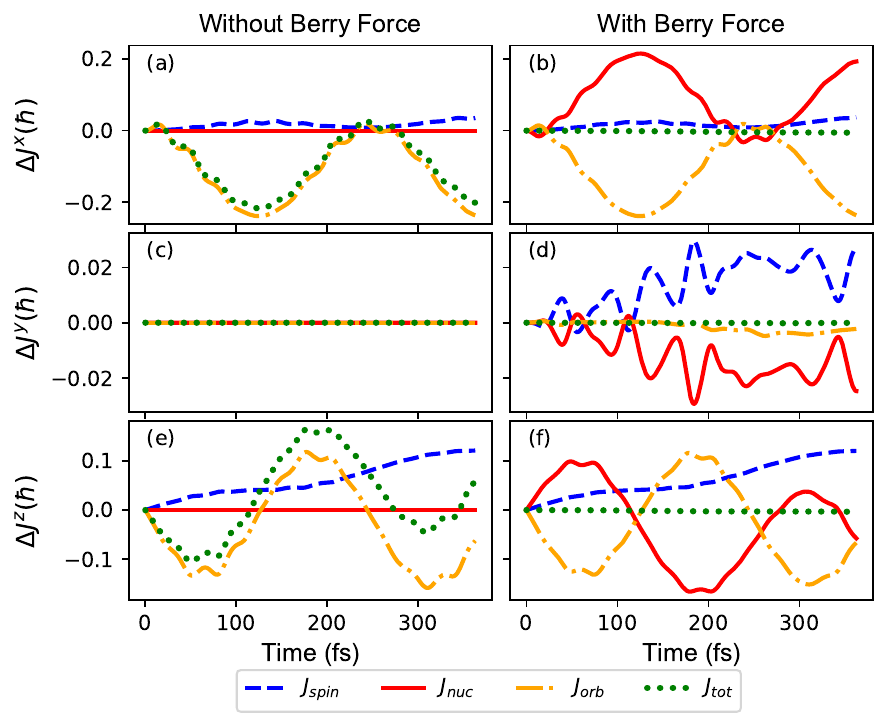}
\caption{ The change in the real-time  angular momentum $\Delta J^\alpha (t) =   J^\alpha (t)-  J^\alpha (0)$ (relative to time zero) according to
an Ehrenfest trajectory of the methoxy radical initiated with a rotational velocity. The nonzero initial values of components of the angular momentum are:  $J^{x}_{\rm orb} = 0.2373 \hbar$, $L^{z}_{\rm orb} = -0.0044 \hbar$,  $J^{y}_{\rm nuc} = 39.28 \hbar$, $J^{x}_{\rm spin} = -0.4999 \hbar$, and $J^{z}_{\rm spin} = 0.0097 \hbar$. The three panels (a),(c),(e) on the left exclude the non-Abelian Berry force; then panels (b),(d) and (f) on the right include the Berry force.  We plot angular momenta along all three Cartesian coordinates, $x$, $y$ and $z$. Without a Berry force, the total angular momentum is not conserved; with a Berry force, the total angular momentum is conserved. 
We find that the nuclear and electronic orbital and spin degrees of freedom transfer angular momentum between each other. Comparing (c) and (d), we observe that the electronic spin in the $y$-direction behaves differently with vs without a Berry force; this change arises because adding the Berry force can change the nuclear motion and yield a reasonably different trajectory. }\label{fig:J_ehren_bf}
\end{figure}

In Fig. \ref{fig:J_ehren_bf}(a)(c)(e) and (b)(d)(f), we compare and contrast the nuclear (red solid line), electronic orbital (orange dashed-dotted line), spin (blue dashed line), and total (green dotted line) angular momentums changes with time obtained without (standard Ehrenfest given in Eqs. (17)-(19) in the main manuscript) and with Berry force (effective Ehrenfest given in Eqs. (68)-(71) in the main manuscript), respectively. As discussed previously,  Ehrenfest dynamics without the Berry force (Fig. \ref{fig:J_ehren_bf}(a)(c)(e)) do not conserve total angular momenutm. Including Berry force does conserve the total angular momentum (Fig. \ref{fig:J_ehren_bf}(b)(d)(f)). Comparing Fig. \ref{fig:J_ehren_bf} (c) and (d), we see that the electronic spin and nuclear angular momentum can behave differently when the Berry force is included. 

